\documentclass[aps,pre,twocolumn,superscriptaddress,showpacs]{revtex4}
\usepackage{epsf}
\usepackage{amsmath,amssymb}
\usepackage{graphicx}
\usepackage{color}

\begin{document}

\title{Complexity of Quantum States and Reversibility of
Quantum Motion}

\date{\today}

\author{Valentin V. Sokolov}
\affiliation{Budker Institute of Nuclear Physics, Novosibirsk,
Russia}
\affiliation{CNISM, CNR-INFM, and Center for Nonlinear and Complex Systems,
Universit\`a degli Studi dell'Insubria, Via Valleggio 11, 22100
Como, Italy}
\author{Oleg V. Zhirov}
\affiliation{Budker Institute of Nuclear Physics, Novosibirsk,
Russia}
\affiliation{CNISM, CNR-INFM, and Center for Nonlinear and Complex Systems,
Universit\`a degli Studi dell'Insubria, Via Valleggio 11, 22100
Como, Italy}
\author{Giuliano Benenti}
\affiliation{CNISM, CNR-INFM, and Center for Nonlinear and Complex Systems,
Universit\`a degli Studi dell'Insubria, Via Valleggio 11, 22100
Como, Italy}
\affiliation{Istituto Nazionale di Fisica Nucleare, Sezione di Milano,
Via Celoria 16, 20133 Milano, Italy}
\author{Giulio Casati}
\affiliation{CNISM, CNR-INFM, and Center for Nonlinear and Complex Systems,
Universit\`a degli Studi dell'Insubria, Via Valleggio 11, 22100
Como, Italy}
\affiliation{Istituto Nazionale di Fisica Nucleare, Sezione di Milano,
Via Celoria 16, 20133 Milano, Italy}
\affiliation{Department of Physics, National University of Singapore,
Singapore 117542, Republic of Singapore}

\begin{abstract}
We present a quantitative analysis of the reversibility
properties of classically chaotic quantum motion. We analyze
the connection between reversibility and the rate at which a
quantum state acquires a more and more complicated structure in
its time evolution. This complexity is characterized by the
number ${\cal M}(t)$ of harmonics of the (initially isotropic,
i.e. ${\cal M}(0)=0$) Wigner function, which are generated
during quantum evolution for the time $t$. We show that, in
contrast to the classical exponential increase, this number can
grow not faster than linearly and then relate this fact with
the degree of reversibility of the quantum motion. To explore
the reversibility we reverse the quantum evolution at some
moment $T$ immediately after applying at this moment an instant
perturbation governed by a strength parameter $\xi$. It follows
that there exists a critical perturbation strength,
$\xi_c\approx \sqrt{2}/{\cal M}(T)$, below which the initial
state is well recovered, whereas reversibility disappears when
$\xi\gtrsim \xi_c(T)$. In the classical limit the number of
harmonics proliferates exponentially with time and the motion
becomes practically irreversible. The above results are
illustrated in the example of the kicked quartic oscillator
model.
\end{abstract}


\maketitle

\section{Introduction}

Strong numerical evidence has been obtained that the quantum
evolution is very stable, in sharp contrast to the extreme
sensitivity to initial conditions and rapid loss of memory
which is the very essence of classical chaos. In computer
simulations the latter effect leads to practical
irreversibility of classically chaotic dynamics. Indeed, even
though the exact equations of motion are reversible, any,
however small, imprecision such as computer round-off errors,
is magnified by the exponential instability of trajectories to
the extent that any memory of the initial conditions is effaced
and reversibility is destroyed. In contrast, almost exact
reversion is observed in numerical simulations of the quantum
motion of classically chaotic systems, even in the regime in
which statistical phenomena such as deterministic diffusion
take place \cite{arrow}.

It is intuitive that the physical reasons of this striking
difference between quantum and classical motion are rooted in
the quantization of the phase space in quantum mechanics. If we
consider classical chaotic evolution (governed by the Liouville
equation) of some phase space distribution, smaller and smaller
scales are explored exponentially fast with time. These fine
details of the density distribution are lost due to finite
accuracy (inevitable coarse-graining) in numerical simulations,
and therefore the reversal of time evolution cannot be carried
out. On the other hand, in quantum mechanics one expects that
the structure of a quantal phase-space distribution, e.g. of
the Wigner function, has resolution limited by the size of the
Planck's cell. While the mean number of Fourier components of
the classical  phase-space distribution grows exponentially in
time for chaotic motion, the number of the components of the
Wigner function at any given time is related to the degree of
excitation of the system (see for example eq. (\ref{PAver_m})
below) and therefore unrestricted exponential growth of this
number is not physical \cite{chirikov,gu,brumer}. This fact
implies substantially simpler phase space structure in the case
of quantum motion as compared with that of classical chaotic
dynamics. We demonstrate below that the mean number of Fourier
Harmonics is a simple  relevant measure of structural
complexity which in turn is related to fundamental properties
such as decoherence and entanglement.

In spite of the above arguments, a rigorous link between the
intuitively expected different degree of reversibility of
quantum and classical motion and the structure developed by the
phase-space distributions during dynamical evolution has never
been established. The purpose of the present paper is to
clarify this problem.

Following the approach developed in \cite{ikeda} we consider
first the forward evolution
\begin{equation}\label{Probe}
{\hat\rho}(t)={\hat U(t)}{\hat\rho}(0){\hat U}^\dagger(t)
\end{equation}
of an initial (generally mixed) state ${\hat\rho}(0)$ up to some time $t=T$. A perturbation ${\hat P}(\xi)$ is then applied at this time, with the perturbation strength $\xi$. For our purposes, it will be sufficient to consider unitary perturbations ${\hat P}(\xi)=e^{-i\xi {\hat V}}$, where ${\hat V}$ is a Hermitian operator. The perturbed state
\begin{equation}
{\hat{\tilde\rho}}(T,\xi)={\hat P}(\xi){\hat\rho}(T){\hat
P}^\dagger(\xi)
\end{equation}
is then evolved backward for the time $T$, thus obtaining the
reversed state
\begin{equation}
\begin{array}{c}
{\hat{\tilde\rho}}(0|T,\xi)={\hat
U}^{\dag}(T){\hat{\tilde\rho}}(T,\xi){\hat U}(T)=\\
\\
{\hat P}(\xi,T){\hat\rho}(0){\hat P}^\dagger(\xi,T),
\end{array}
\end{equation}
where ${\hat P}(\xi,T)\equiv e^{-i\xi {\hat V}(T)}$, with
${\hat V}(T)\equiv {\hat U}^\dagger(T){\hat V}{\hat U}(T)$
being the Heisenberg evolution of the perturbation during the
time $T$.

Finally, we investigate the distance between the reversed
${\hat{\tilde\rho}}(0|T,\xi)$ and the initial ${\hat\rho}(0)$
state, as measured by the Peres fidelity \cite{peres}
\begin{equation}\label{PFid}
\begin{array}{c}
{\displaystyle
F(\xi;T)=\frac{{\rm
Tr}[{\hat{\tilde\rho}}(0|T,\xi){\hat\rho(0)}]} {{\rm
Tr}[{\hat\rho}^2(0)]}}\\
{\displaystyle
=\left.\frac{{\rm Tr}[{\hat{\tilde\rho}}(t,\xi){\hat\rho(t)}]} {{\rm
Tr}[{\hat\rho}^2(t)]}\right|_{t=T}=F\left.(\xi;t)\right|_{t=T}\,.}
\end{array}
\end{equation}
This quantity is bounded in the interval $[0,1]$ and the
distance between the initial and the time-reversed state is
small when $F(\xi;T)$ is close to one~\cite{fidelitynote}. In
particular, $F(\xi;T)=1$ when the two states coincide. The
second line in Eq.~(\ref{PFid}) is a consequence of the unitary
time evolution and will allow us to relate the distance between
the initial and the reversed state to the complexity of the
state $\hat{\rho}(t)$ at the reversal time $t=T$. This relation
will play the key role in our further analysis.

As we have already mentioned above, we characterize the
complexity of a quantum state by the structure of its Wigner
function. The basic idea here is that a quantum state is
complex if this function has a rich phase space structure,
which can be naturally measured by the number ${\cal M}(t)$ of
its Fourier harmonics. We will show that the fidelity
$F(\xi;T)$ is a decreasing function of the complexity ${\cal
M}(T)$ of the state ${\hat\rho}(T)$. We will prove further
that, after the Ehrenfest time scale, namely after a time
logarithmically short in the effective Planck constant of the
system, the number ${\cal M}(t)$ of harmonics increases
notfaster than linearly with time. On the other hand, in
classical chaotic dynamics the number of harmonics ${\cal
M}_c(t)$ of the classical phase-space distribution function
grows exponentially in time. We will then ascertain that the
initial state is well recovered as long as the perturbation
strength $\xi$ is much smaller than a critical perturbation
strength $\xi_c(T)\sim 1/{\cal M}(T)$. Therefore, $\xi_c(T)$
drops exponentially with $T$ in the classical case and not
faster than linearly for quantum evolution. This fact explains
the much weaker sensitivity of quantum dynamics to
perturbations as compared to the classical chaotic motion.

The paper is organized as follows. In Sec.~II, we review the
main concepts and definitions of quantum mechanics in phase
space relevant for our work. In Sec.~III, we discuss the kicked
quartic oscillator model, used in the remaining part of the
paper as a test bed to illustrate the stability properties of
classically chaotic quantum motion. In Sec~IV, the evolution in
time of the harmonics of the Wigner function is studied in
detail. In particular, the complexity of the quantum state
${\hat\rho}(t)$ is quantified by looking at the sensitivity of
the system to an infinitesimal perturbation. In Sec.~V, the
degree of reversibility of quantum motion, as measured by the
Peres fidelity, is related to the number of harmonics developed
by dynamics at the reversal time $t=T$. A more detailed study
of the reversibility properties of motion is carried out in
Sec.~VI, where we investigate the properties of the
time-reversed state by studying the harmonics of its Wigner
function. Finally, the main results of our paper are summarized
in Sec.~VI.

\section{Quantum dynamics in the phase space}

The phase-space representation of quantum mechanics is a very
enlightening approach which allows a direct comparison with
classical mechanics~\cite{gu,brumer}. In this section, we
briefly review the main aspects of the phase-space approach
which are relevant for our work.

\subsection{The Wigner Function}

Let us consider a nonlinear system whose dynamics is governed
by the Hamiltonian operator ${\hat H}\equiv H({\hat
a}^{\dag},{\hat a};t)=H^{(0)}({\hat n}={\hat a}^{\dag}{\hat
a})+H^{(1)}({\hat a}^{\dag},{\hat a};t)$ with the
time-independent unperturbed part ${\hat H}^{(0)}$ which has a
discrete energy spectrum bounded from below so that we can
assume that all its eigenvalues $E_n^{(0)}\geqslant 0$. Here
${\hat a}^{\dag}, {\hat a}$ are the bosonic, $[{\hat a},{\hat
a}^{\dag}]=1\,$, creation-annihilation operators. We will use
the method of c-number $\alpha$-phase space borrowed from the
quantum optics (see for example \cite{Glauber63,Agarwal70}).
This method is equally suitable for analyzing both the quantum
and classical evolutions. It is, basically, built upon the
basis of the coherent states $|\alpha\rangle$. The latter are
defined by the eigenvalue problem ${\hat
a}|\alpha\rangle=\frac{\alpha}{\sqrt \hbar}|\alpha\rangle$,
where $\alpha$ is a complex variable independent of $\hbar$. An
arbitrary coherent state is obtained from the ground state,
$|\alpha\rangle={\hat
D}\left(\frac{\alpha}{\sqrt\hbar}\right)|0\rangle$, with the
help of the unitary displacement operator ${\hat
D}\left(\lambda\right)=\exp(\lambda\,{\hat
a}^{\dag}-\lambda^*{\hat a})$.

Using the displacement operator ${\hat D}$ as a kernel, one can
represent any operator function ${\hat G}(t)\equiv G({\hat
a}^{\dag},{\hat a};t)$  in the form of the operator Fourier
transformation \cite{Agarwal70}
\begin{equation}\label{OpFourTr}
G({\hat a}^{\dag},{\hat a};t)=\frac{1}{\pi}\int
d^2\eta\,{\tilde G}(\eta^*,\eta;t)\,{\hat D}(\eta)\,,
\end{equation}
where ${\tilde G}(\eta^*,\eta;t)$ is a numerical function of
two independent complex variables $\eta^*, \eta$ and the
integration runs over the complex $\eta$-plane. The inverse
transformation
\begin{equation}\label{InvFourTr}
\tilde{G}(\eta^*,\eta;t)=Tr\left[G(\hat{a}^{\dag},\hat{a};t)\,
\hat{D}^{\dag}(\eta)\right]
\end{equation}
is immediately obtained with the help of the orthogonality
condition for the displacement operators,
\begin{equation}\label{OrthD}
\frac{1}{\pi}\,Tr\left[\hat{D}^{\dag}(\eta')\,\hat{D}(\eta)\right]=
\delta^{(2)}(\eta'-\eta)\,.
\end{equation}

By using transformation (\ref{OpFourTr}), the standard
quantum-mechanical formula $\langle
Q\rangle=Tr\left[{\hat\rho}(t)\,{\hat Q}\right]$ for the mean
expectation value $\langle Q\rangle$ of a dynamical variable
$Q$ [$Q$ is represented by the operator ${\hat
Q}=Q(\hat{a}^{\dag},\hat{a})$] in a generally mixed state
${\hat\rho}(t)=\rho({\hat a}^{\dag},{\hat a};t)$ can be written
in a way formally equivalent to the classical phase space
average:
\begin{equation}\label{PhAver}
\begin{array}{c}
\langle Q\rangle=\frac{1}{\pi}\int
d^2\eta\,{\tilde\rho}(\eta^*,\eta;t)
\,{\tilde Q}(-\eta^*,-\eta)=\\
\,\,\,\,\,\,\\
\int d^2\alpha W(\alpha^*,\alpha;t)\,Q(\alpha^*,\alpha)\,.
\end{array}
\end{equation}
The final form is readily obtained after defining the Wigner
function $W(\alpha^\star,\alpha;t)$ and
$Q(\alpha^\star,\alpha)$ as c-number Fourier transformations of
$\tilde{\rho}(\eta^\star,\eta;t)$ and
$\tilde{Q}(\eta^\star,\eta)$, respectively. The Wigner function
in the $\alpha$-phase plane is connected to the density
operator $\hat{\rho}(t)$ as
\begin{equation}\label{Wfunc}
\begin{array}{c}
W(\alpha^*,\alpha;t)=
\frac{1}{\pi^2\hbar}\int d^2\eta\,
\exp\left(\eta\frac{\alpha^\star}{\sqrt\hbar}-
\eta^\star\frac{\alpha}{\sqrt\hbar}\right)
\tilde{\rho}(\eta^\star,\eta;t)=\\
\frac{1}{\pi^2\hbar}\int d^2\eta\,
\exp\left(\eta^*\frac{\alpha}{\sqrt\hbar}-
\eta\frac{\alpha^*}{\sqrt\hbar}\right) Tr\left[{\hat\rho(t)}\,{\hat
D(\eta)}\right]
\end{array}
\end{equation}
(this corresponds to the Weyl's ordering of the
creation-annihilation operators). Similarly,
\begin{equation}\label{Qfunc}
\begin{array}{c}
Q(\alpha^*,\alpha)=
\frac{1}{\pi}\int d^2\eta\,
\exp\left(\eta\frac{\alpha^\star}{\sqrt\hbar}-
\eta^\star\frac{\alpha}{\sqrt\hbar}\right)
\tilde{Q}(\eta^\star,\eta)=\\
\frac{1}{\pi}\int d^2\eta\,
\exp\left(\eta^*\frac{\alpha}{\sqrt\hbar}-
\eta\frac{\alpha^*}{\sqrt\hbar}\right) Tr\left[{\hat Q}\,{\hat
D(\eta)}\right]\,.\\
\end{array}
\end{equation}

It follows from the definition (\ref{Wfunc}) that the Wigner
function is normalized to unity:
\begin{equation}\label{N}
\int d^2\alpha W(\alpha^*,\alpha;t)=Tr{\hat\rho(t)}=1\,.
\end{equation}
The Wigner function is real but, unlike its classical
counterpart, is not in general positive definite.

With the help of the Wigner function the Peres fidelity
(\ref{PFid}) can be expressed as
\begin{equation}\label{FidW}
\begin{array}{c}
{\displaystyle
F(\xi;T)=\frac{\int d^2\alpha\,W\left(\alpha^*,\alpha;0\right)
\tilde{W}\left(\alpha^*,\alpha;0\big|T,\xi\right)}
{\int d^2\alpha\,W^2\left(\alpha^*,\alpha;0\right)}=}\\
{\displaystyle
\frac{\int d^2\alpha\,W\left(\alpha^*,\alpha;T\right)
\tilde{W}\left(\alpha^*,\alpha;T,\xi\right)} {\int
d^2\alpha\,W^2\left(\alpha^*,\alpha;T\right)}\,.}
\end{array}
\end{equation}
The important advantage of this representation is that it
remains valid in the classical case when the Wigner function
reduces to the classical distribution function,
$W^{(c)}(\alpha^*,\alpha;t)\,$.

\subsection{Harmonics of the Wigner Function}

We define the harmonic's amplitudes $W_m(I;t)$ of the Wigner
function by the Fourier expansion
\begin{equation}\label{Four}
W(\alpha^*,\alpha;t)=\frac{1}{\pi}\,\sum_{m=-\infty}^{\infty}W_m(I;t)\,
e^{i m\theta}\,,
\end{equation}
where $\alpha=\sqrt{I}\,e^{-i\theta}\,.$ Then the normalization
condition simply implies that
$\int_0^{\infty}dI\,W_0(I;t)=1\,.$ There are no restrictions on
$W_m$ when $m\neq 0$.

The amplitudes $W_m$ can be expressed in terms of the matrix
elements $\langle n+m | \hat{\rho}|n\rangle$ along $m$-th
subdiagonal of the density matrix. Indeed, using the well-known
\cite{Schwinger53} matrix elements of the displacement operator
in the basis of the eigenvectors $|n\rangle$ of the unperturbed
Hamiltonian ${\hat H}^{(0)}$,
\begin{equation}\label{D}
\langle n+m |{\hat
D}(\eta)|n\rangle={\sqrt\frac{n!}{(n+m)!}}\,\eta^m\,
e^{-\frac{1}{2}|\eta|^2}L_n^m (|\eta|^2)\,,
\end{equation}
($n,m\geq 0$) where $L_n^m (x)$ is the Laguerre polynomial, the
$\eta$-integration in the second line of Eq.~(\ref{Wfunc}) can
be carried out explicitly. We obtain finally
\begin{equation}\label{Harm}
\begin{array}{c}
W_m(I;t)=\frac{2}{\hbar}\,e^{-\frac{2}{\hbar}I}\sum_{n=0}^{\infty}(-1)^n
\sqrt{\frac{n!}{(n+m)!}}\times\\
\left(4I/\hbar\right)^{\frac{m}{2}}
L_n^m\left(4I/\hbar\right)\langle n+m |{\hat\rho}(t)|n\rangle
\end{array}
\end{equation}
when $m\geq 0$ and $W_{-m}=W_{m\geq 0}^\star$.

With the help of the orthogonality condition for the Laguerre
polynomials Eq.~(\ref{Harm}) can be inverted, thus obtaining
\begin{equation}\label{RHarm}
\begin{array}{c}
\langle n+m\big|{\hat{\rho}}(t)\big|n\rangle=(-1)^n\,
2\sqrt{\frac{n!}{(n+m)!}}\times\\
\int_0^{\infty}\!dI\,e^{-2\frac{I}{\hbar}}\,
\left(4I/\hbar\right)^{\frac{m}{2}}
L_n^m\left(4I/\hbar\right)\, W_m\left(I;t\right)\,.
\end{array}
\end{equation}

\subsection{Time - Evolution of the Wigner Function}

The Wigner function satisfies the evolution equation
\begin{equation}\label{Leq}
i\frac{\partial}{\partial t}\,W(\alpha^*,\alpha;t)= {\cal\hat
L}_q\,W(\alpha^*,\alpha;t).
\end{equation}
Here ${\cal\hat L}_q is $ the Hermitian ``quantum Liouville
operator'' ${\cal\hat L}_q$ whose explicit form is obtained by
mapping the standard equation
\begin{equation}\label{DenM_Eq}
i\frac{\partial}{\partial t}{\hat\rho}(t)=\frac{1}{\hbar}[{\hat
H},{\hat\rho}(t)]
\end{equation}
onto the $\alpha$-phase space. Supposing that the Hamiltonian
can be presented as a Hermitian sum (finite or infinite) of
products of the creation-annihilation operators and using then
the definitions of the phase space images, Eqs.~(\ref{Wfunc}),
(\ref{Qfunc}), it is possible to show that (see
Ref.~\cite{Agarwal70})
\begin{equation}\label{Liouv_Ham}
\begin{array}{c}
{\cal\hat L}_q= \frac{1}{\hbar}\left[{\cal H}\left(\alpha^*-
\frac{\hbar}{2}\frac{\vec{\partial}}{\partial\alpha}, \alpha+
\frac{\hbar}{2}\frac{\vec{\partial}}{\partial\alpha^*}\right)-\right.\\
\left.{\cal H}\left(\alpha^*+
\frac{\hbar}{2}\frac{\vec{\partial}}{\partial\alpha}, \alpha-
\frac{\hbar}{2}\frac{\vec{\partial}}{\partial\alpha^*}\right)\right],
\end{array}
\end{equation}
where ${\cal H}(\alpha^*,\alpha)$ is the phase-space image of
the system's Hamiltonian. The arrows above the derivatives mean
that they act only on the arguments of the Wigner function but
ignore the $(\alpha^*,\alpha)$-dependence of the phase-space
operator ${\cal H}$ itself. In the classical limit $\hbar=0$
the operator ${\cal\hat L}_q$ has the standard classical form
\begin{equation}\label{Liouv_c}
\hat{{\cal L}}_c=\frac{\partial
H_c(\alpha^*,\alpha;t)}{\partial\alpha}\,\frac{\partial}{\partial\alpha^*}-
\frac{\partial
H_c(\alpha^*,\alpha;t)}{\partial\alpha^*}\,\frac{\partial}{\partial\alpha},
\end{equation}
where the classical Hamiltonian function coincides with the
diagonal matrix element,
$H_c(\alpha^*,\alpha;t)=\langle\alpha|{\hat H}^{(N)}({\hat
a}^{\dag},{\hat a})|\alpha\rangle$ of the normal form ${\hat
H}^{(N)}$ of the quantum Hamiltonian operator. In other words,
this function is obtained from the quantum Hamiltonian by
substituting ${\hat a}\rightarrow
\alpha/\sqrt{\hbar}\,,\,\,{\hat
a}^{\dag}\rightarrow\alpha^*/\sqrt{\hbar}\,.$

The outlined phase-space approach is quite general and can be
readily extended  \cite{Glauber63, Agarwal70} to systems with
arbitrary number of degrees of freedom and is applicable to any
system whose Hamiltonian can be expressed in terms of a set of
the bosonic creation-annihilation operators~\footnote{For
instance, this approach can be used for systems like the kicked
rotor model, with the motion confined to a ring geometry}. In
this approach, the classical distribution function as well as
its quantum counterpart are described in terms of the same
phase-space variables, thus allowing a straightforward
comparison of the two dynamics \cite{gu, brumer}.

\section{The Model}

In order to discuss the reversibility/complexity properties of
quantum motion, we consider, as an illustrative example, the
kicked quartic oscillator model, described by the Hamiltonian
\cite{Berman78,chirikov,Sokolov84}
\begin{equation}\label{Ham}
{\hat H}=H({\hat a}^{\dag},{\hat a})=\hbar\,\omega_0 {\hat
n}+\hbar^2\,{\hat n}^2-\sqrt{\hbar}\,g(t)({\hat a}+{\hat
a}^{\dag}),
\end{equation}
where $g(t)=g_0\sum_s\delta(t-s)$, and ${\hat n}={\hat
a}^{\dag}{\hat a}$, $[{\hat a},{\hat a}^{\dag}]=1$. In our
units, the time and parameters $\hbar, \omega_0$ as well as the
strength of the driving force are dimensionless. The period of
the driving force $g(t)$ is set to one. The corresponding
classical Hamiltonian function $H_c$ can be expressed in terms
of complex canonical variables $\alpha, i{\alpha}^*$ which are
related to the classical action-angle variables $I, \theta$ via
$\alpha=\sqrt{I}e^{-i\theta},\alpha^*=\sqrt{I}e^{i\theta}$. It
reads
\begin{equation}
\label{Hamc}
H_c=\omega_0|\alpha|^2+|\alpha|^4-g(t)(\alpha^*+ \alpha)\,.
\end{equation}
Detailed analytical semiclassical analysis of the quantum
motion of the model (\ref{Ham}) has been presented in
\cite{Sokolov84,Sokolov07}.

The quantum Liouville operator for the model (\ref{Ham}) can be
derived from Eq.~(\ref{Liouv_Ham}) and reads
\begin{equation}\label{ML_q}
\begin{array}{c}
{\displaystyle
{\cal\hat L}_q={\cal\hat L}_q^{(0)}+
{\cal\hat L}^{(kick)}},\\
\displaystyle{
{\cal\hat L}_q^{(0)}=
\left(\omega_0-\hbar-\frac{1}{2}\hbar^2
\frac{\partial^2}{\partial\alpha^*\partial\alpha}+2|\alpha|^2\right)
\left(\alpha^*\frac{\partial}{\partial\alpha^*}-
\alpha\frac{\partial}{\partial\alpha}\right)},
\\
{\displaystyle
{\cal\hat L}^{(kick)}=
-g(t)\left(\frac{\partial}{\partial\alpha^*}-
\frac{\partial}{\partial\alpha}\right)}.
\end{array}
\end{equation}
The essential difference between this operator and the
corresponding classical Liouville operator ${\cal\hat L}_c$ is
the presence in Eq.~(\ref{ML_q}) of a term proportional to
$\hbar^2$ which contains a second derivative over the phase
space variables. This term drastically changes the spectrum of
the unperturbed part ${\cal\hat L}_q^{(0)}$ of the Liouville
operator and thereby modifies the evolution of the Wigner
distribution function $W$ with respect to the classical
distribution function $W^c$. Indeed, while in the classical
limit $\hbar=0$ the factor
$\omega_0+2|\alpha|^2=\omega_0+2I=\omega_c(I)$ is the
continuous frequency of the classical quartic oscillator, the
quantum operator ${\hat K}\equiv
-\frac{1}{2}\hbar^2\frac{\partial^2}{\partial\alpha^*\partial\alpha}+
2|\alpha|^2$
has a discrete spectrum: considering the real and imaginary
parts $(\alpha_1,\alpha_2)$ of the variable $\alpha$ as
cartesian coordinates, we obtain
\begin{equation}
{\hat K}=-\frac{\hbar^2}{8}\left( \frac{\partial^2}{\partial
\alpha_1^2}+ \frac{\partial^2}{\partial \alpha_2^2}\right)
+2(\alpha_1^2+\alpha_2^2).
\end{equation}
Therefore, the operator $\hat{K}$ is formally equivalent to the
Hamiltonian operator of a two-dimensional isotropic oscillator,
with the frequency $\nu=1$ and the mass $\mu=4$. After
introducing the standard annihilation-creation operators
\begin{equation}\label{LinPol}
\begin{array}{c}
{\hat A}_{1,2}=\sqrt{\frac{2}{\hbar}}\,\alpha_{1,2}+
\frac{1}{\sqrt{8\hbar}}\,\frac{\partial}{\partial\alpha_{1,2}},\\
{\hat A}_{1,2}^{\dag}=\sqrt{\frac{2}{\hbar}}\,\alpha_{1,2}-
\frac{1}{\sqrt{8\hbar}}\,\frac{\partial}{\partial\alpha_{1,2}},
\end{array}
\end{equation}
this operator transforms into ${\hat K}=\hbar{\hat N}$, where
${\hat N}= {\hat N}_1+{\hat N}_2={A}_{1}^{\dag}{{\hat
A}_{1}}+{A}_{2}^{\dag} {{\hat A}_{2}}$ is the operator
representing the total number of fictitious quanta linearly
polarized in the $\alpha$-plane. The operator
\begin{equation}\label{Moperator}
\hat{M}\equiv\left(\alpha^*\frac{\partial}{\partial\alpha^*}-
\alpha\frac{\partial}{\partial\alpha}\right)=
-i\frac{\partial}{\partial \theta}
\end{equation}
is proportional to the angular momentum projection operator
along the axis orthogonal to the $\alpha$-plane and has the
discrete eigenvalue spectrum $m=0,\pm 1,\pm 2,...$. We can
therefore conclude that the spectrum of the operator
\begin{equation}\label{SpL1}
{\cal\hat L}_q^{(0)}= \left(\omega_0+\hbar\,{\hat
N}\right){\hat M}
\end{equation}
is also discrete with eigenvalues
$(\lambda_0)_{n,m}=(\omega_0+\hbar n) m$. Since the operators
${\hat N}$ and ${\hat M}$ commute, operator (\ref{SpL1}) is
Hermitian.

The two linearly polarized quanta introduced above are coupled
to each other because the angular momentum operator $\hbar{\hat
M}=\frac{1}{i}\left({\hat A}_1^{\dag}{\hat A}_2-{\hat
A}_2^{\dag}{\hat A}_1\right)$ is not diagonal in the chosen
representation. However both the operators $\hbar{\hat N}$ and
$\hbar{\hat M}$ are simultaneously diagonalized after
introducing the operators
\begin{equation}\label{caop}
{\hat A}_+=\frac{1}{\sqrt{\hbar}}\,\alpha +
\frac{\sqrt{\hbar}}{2}\frac{\partial}{\partial\alpha^*}\,,
\quad {\hat A}_-=\frac{1}{\sqrt{\hbar}}\,\alpha^*+
\frac{\sqrt{\hbar}}{2}\frac{\partial}{\partial\alpha}\,,
\end{equation}
which describe the circularly polarized quanta. In the new representation
\begin{equation}
{\hat N}={\hat N}_+ +{\hat N}_-,\,\,\,{\hat M}=({\hat N}_+
-{\hat N}_-),
\end{equation}
where ${\hat N}_{\pm}={\hat A}_{\pm}^{\dag}\,{\hat A}_{\pm}$
are the operators representing the numbers $n_{\pm}$ of
circularly polarized quanta. These new quanta are decoupled:
\begin{equation}\label{L^0cir}
\begin{array}{c}
{\cal\hat L}_q^{(0)}=\left(\omega_0+\hbar\,{\hat N}\right){\hat
M}={({\cal\hat L}_0)}_+ -{({\cal\hat L}_0)}_-,\\
{({\cal\hat L}_0)}_{\pm}= \omega_0\,{\hat N}_{\pm}+ \hbar {\hat
N}_{\pm}^2\,.
\end{array}
\end{equation}
The eigenvalues
$(\lambda_0)_{n_+,n_-}=\frac{E^{(0)}_{n_+}-E^{(0)}_{n_-}}
{\hbar}$ of the operator (\ref{L^0cir}) are determined by the
distances between the unperturbed energy levels
$E_{n_\pm}^{(0)}=\hbar\omega_0\,n_{\pm}+\hbar^2\,n_{\pm}^2$
corresponding to the excitation numbers $n_{\pm}=0,1,2, ...$.
The representation (\ref{L^0cir}) is the most convenient for
numerical simulations.

The driving perturbation also decouples in the chosen representation:
\begin{equation}\label{Kickcirc}
\begin{array}{c}
{\cal\hat L}^{(kick)}={\cal\hat L}_+^{(kick)}
-{\cal\hat L}_-^{(kick)}\,,\\
{\cal\hat L}_{\pm}^{(kick)}= -g(t)\frac{1}{\sqrt\hbar}\left({\hat
A}_{\pm}+ {\hat A}_{\pm}^{\dag}\right)\,.
\end{array}
\end{equation}
It then immediately follows that the one-period unitary
evolution operator ${\cal\hat F}$ for the Wigner function gets
factorized as
\begin{equation}\label{Ffak}
\begin{array}{c}
{\cal\hat F}={\cal\hat F}_+\,{\cal\hat F}_-^{\dag}\,, \quad
{\cal\hat F}_{\pm}=e^{-i{({\cal\hat L}_0)}_{\pm}}\, {\hat
D}_{\pm}\left(i\frac{g_0}{\sqrt\hbar}\right)\,,\\
{\hat D}_{\pm}\left(i\frac{g_0}{\sqrt\hbar}\right)=
\exp\left(i\frac{g_0}{\sqrt\hbar} \left({\hat A}_{\pm}+{\hat
A}_{\pm}^{\dag}\right)\right)\,.
\end{array}
\end{equation}

The complete set of the eigenvectors
$|n_+\,n_-\rangle=|n_+\rangle|n_- \rangle$ of the unperturbed
operator ${\cal\hat L}_q^{(0)}$ constitutes the excitation
number reference basis for the density matrix,
\begin{equation}\label{DMat}
\begin{array}{c}
{\hat\rho(t)}=\sum_{n_+,n_-}|n_+\rangle\rho(n_+,n_-;t)\langle
n_-|\,;\\
\rho(n_+,n_-;t)=\langle n_+|{\hat\rho(t)}|n_-\rangle\,.
\end{array}
\end{equation}
The evolution from time $t$ to time $t+1$ reads
\begin{equation}\label{1kick}
\begin{array}{c}
\rho(n_+,n_-;t+1)=\langle n_+|{\hat U}_1{\hat\rho(t)}{\hat U}_1^{\dag}|n_
-\rangle=\\
\sum_{n_+',n_-'}\langle n_+|{\cal\hat F}_+|n_+'\rangle
\rho(n_+',n_-';t)\langle n_-'|{\cal\hat F}_-|n_-\rangle^*\,.
\end{array}
\end{equation}
Here
\begin{equation}\label{Flop}
\begin{array}{c}
{\hat U}_1\equiv {\hat U}(t=1)=e^{-\frac{i}{\hbar}{\hat
H}^{(0)}}\, {\hat D}\left(i\frac{g_0}{\sqrt\hbar}\right)\\
= e^{-i\left(\omega_0 {\hat n}+\hbar {\hat n}^2\right)}\,
e^{i\frac{g_0}{\sqrt{\hbar}}\left({\hat a}+{\hat
a}^{\dag}\right)}
\end{array}
\end{equation}
is the (Floquet) operator for our model, namely the one-period
unitary evolution induced by the Hamiltonian (\ref{Ham}).

Our numerical simulations are based on the combined application
of Eqs.~(\ref{PFid}), (\ref{Wfunc}), and (\ref{1kick}). We
calculate numerically the truncated Floquet matrix ${\hat F}_{n
n'}$ in the excitation number representation. Two different
strategies have been used.

The first approach is based on Eq.~(\ref{D}), which relates
matrix elements of the kick operator ${\hat
D}\left(i\frac{g_0}{\sqrt\hbar}\right)$ to the Laguerre
polynomials. The free rotation operator
$e^{-\frac{i}{\hbar}{\hat H}^{(0)}}$ is diagonal and its
calculation is trivial. Then the product of the matrices
$e^{-\frac{i}{\hbar}{\hat H}^{(0)}}$ and $\hat{D}$ is truncated
to a square matrix of a finite size $N$. The main disadvantage
of this approach is the violation of unitarity of the Floquet
operator: its norm is not conserved when the size of a quantum
state becomes of the order or larger than $N$.

The second approach is based on the truncation of the Hermitian
matrix ${\hat X}_{n n'}=\langle
n|\frac{g_0}{\sqrt{\hbar}}\left({\hat a} +{\hat
a}^{\dag}\right)|n'\rangle$. Then we numerically diagonalize
this matrix, ${\hat X}= {\hat v}{\hat X}_d {\hat v}^{\dag}$,
with the aid of a unitary matrix ${\hat v}$. In such a way we
obtain the truncated kick matrix ${\hat
D}(i\frac{g_0}{\sqrt{\hbar}})={\hat v}\exp(i {\hat X}_d){\hat
v}^{\dag}$ in the excitation number basis. Multiplying it by
the, diagonal in this basis, free rotation matrix
$\exp\left(-\frac{i}{\hbar}{\hat H}^{(0)}\right)$ (truncated to
the same size $N$) and finally diagonalizing the obtained
unitary matrix, we arrive at a truncated approximation ${\hat
F}={\hat V}{\hat F}_d {\hat V}^{\dag}$ of the Floquet operator
(\ref{Flop}) in the excitation number representation. The price
paid is the artificial boundary condition at $n=N$ which
influences the evolution when the size of a quantum state
becomes close to that of the truncated Floquet operator. Even
though the final diagonalization problem is, by itself, rather
time-consuming the great advantage of this approach is that the
computation time does not depend on the considered duration of
the evolution. This is due to the fact that
$\hat{U}(t)=\hat{F}^t={\hat V}{\hat F}_d^t {\hat V}^{\dag}$ and
therefore, independently of $t$, it is sufficient to multiply
$3$ matrices to construct $\hat{U}(t)$.

\section{Harmonics Dynamics}

In this section, we study the time evolution of the harmonics
of the Wigner function for the kicked quartic oscillator model.
The Wigner function of the initial state is taken to be
isotropic, so that only the zero angular harmonic $W_0$ of the
Wigner function is different from zero. In the course of
dynamical evolution the Wigner function becomes more and more
structured and this reflects in the excitation of a growing
number of harmonics. The complexity of the Wigner function at
time $t$ is measured by its sensitivity to an infinitesimal
perturbation.

\subsection{Initial Condition}

We choose the initial state to be an isotropic mixture of coherent states:
\begin{equation}\label{IsInMix}
{\hat\rho}(0)= \int d^2\overset{\circ}\alpha\, {\cal
P}(|\overset{\circ}\alpha|^2)|\overset{\circ}\alpha\rangle
\langle\overset{\circ}\alpha|=\sum_{n=0}^{\infty}\,
{\rho}_n\,|n\rangle\langle n|,
\end{equation}
where
\begin{equation}\label{InRho_n}
{\rho}_n=\frac{\pi}{n!}\int_0^{\infty} d\overset{\circ}I\,
{\cal P}(\overset{\circ}I)\, e^{-\overset{\circ}I/\hbar}
\left(\overset{\circ}I/\hbar\right)^n, \,\,\,\,
\overset{\circ}I= |\overset{\circ}\alpha|^2\,.
\end{equation}
Here and in the following, a circle above a dynamical variable
denotes its value at the time $t=0$. To simplify further
analytical considerations we suppose a Poissonian initial
distribution ${\rho}_{n n}\equiv
{\rho}_n=\frac{\hbar}{\Delta+\hbar}\,
\left(\frac{\Delta}{\Delta+\hbar}\right)^n$ in the excitation
number space, which implies the exponential form ${\cal
P}(\overset{\circ}I)
=\frac{1}{\pi\Delta}\,e^{-\overset{\circ}I/\Delta}$ of the
distribution in the space  of coherent states and,
correspondingly, the isotropic Gaussian initial Wigner function
\begin{equation}\label{InWfunc}
W(\alpha^*,\alpha;0)=\frac{1}{\Delta+\hbar/2}\,
e^{-\frac{|\alpha|^2}{\Delta+\hbar/2}}\,.
\end{equation}
In the particular case of the pure ground state, $\Delta=0$,
the Wigner function occupies the minimal quantum cell with the
area $\hbar/2\,.$ It is worth noting that this area would be
twice as much in the case of the normal ordering of the
creation annihilation operators (Husimi function)
\cite{Glauber63, Agarwal70}.

\begin{figure}
\includegraphics[height=7.cm,angle=90]{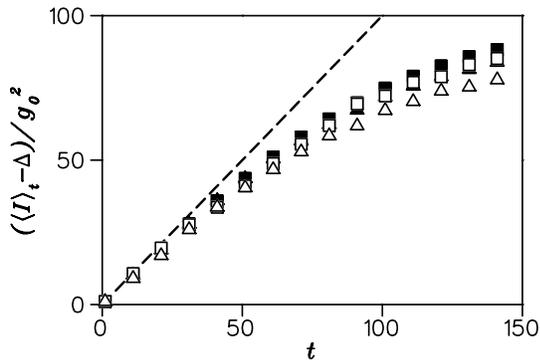}
\caption{Mean value $(\langle I\rangle_t-\Delta)/ g_0^2$ as a
function of time $t$. Squares and triangles correspond to
$(\hbar,g_0)$=$(1,2)$ and $(2,3)$, full and empty symbols to
$\Delta=0$ and $50$. The straight dashed line coresponds to the
classical diffusion law $\langle I \rangle_t=\Delta+g_0^2 t$.}
\label{fig:avn}
\end{figure}

For the initial conditions (\ref{IsInMix}) the classical
dynamics of the model (\ref{Ham}) becomes chaotic when the kick
strength parameter $g_0$ exceeds a critical value
$g_{0,c}\approx 1$. The angular phase correlations decay
exponentially (we have checked this fact numerically) and the
mean action grows diffusively with the diffusion coefficient
$D= g_0^2.$ Our numerical data presented in Fig.~\ref{fig:avn}
demonstrate the corresponding "quantum diffusion" phenomenon
described in the following subsection: $\langle
I\rangle_t=\Delta+g_0^2\,t$ as in the classical case, until a
time $t^\star$ after which the quantum to classical
correspondence breaks down \cite{chirikov,Berman78}.

\subsection{Zero-Harmonic Evolution and Quantum Diffusion}

\begin{figure}
\includegraphics[height=7.cm,angle=90]{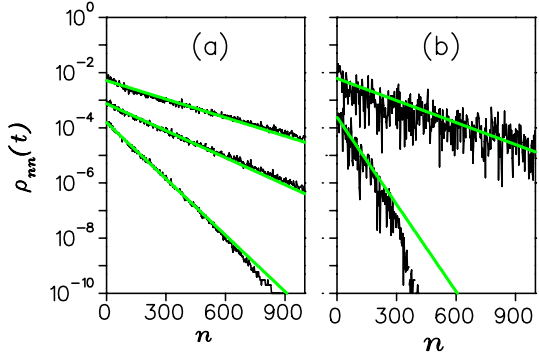}
\caption{(color online) Distribution of the diagonal elements
$\rho_{nn}(t)$ of the density matrix versus $n$, at $\hbar=1$,
$g_0=2$. \textit{Left panel} (a): mixed state ($\Delta=25$) and, from bottom
to top, $t=10, 30$, and $50$ (these curves are scaled by
factors 0.01, 0.1 and 1, respectively). \textit{Right panel} (b): pure state
($\Delta=0$), $t=10$ (bottom, scaled by a factor $0.01$) and
$t=50$ (top). Straight lines show exponential fits,
corresponding to the coarse-grained distribution
(\ref{DiagAnz}).} \label{fig:rhodiag}
\end{figure}

According to Eq.~(\ref{Qfunc}), the zero amplitude $W_0(I;t)$
is determined as
\begin{equation}\label{W_0}
W_0(I;t)=\frac{2}{\hbar}\,e^{-\frac{2}{\hbar}I}\sum_{n=0}^{\infty}(-1)^n
L_n\left(4I/\hbar\right)\langle n|{\hat\rho}(t)|n\rangle
\end{equation}
by the diagonal matrix elements of the density operator. Our
numerical simulations (see Fig.~\ref{fig:rhodiag}) show that
these matrix elements decay along the main diagonal on average
exponentially at any moment $t$ after a short initial interval
(see below). Neglecting fluctuations we therefore assume that
\begin{equation}\label{DiagAnz}
\begin{array}{c}
\rho_{nn}(t)=\langle n|{\hat\rho}(t)|n\rangle= \int
d^2\overset{\circ}\alpha\, {\cal
P}(|\overset{\circ}\alpha|^2)\Big|\langle n\big|{\hat
U}(t)\big|\overset{\circ}\alpha\rangle\Big|^2=\\
\sum_{n'=0}^{\infty}\rho_{n'} \Big|\langle n\big|{\hat
U}(t)\big|n'\rangle\Big|^2
=\left[1-e^{-\nu(t)}\right]\,e^{-\nu(t)n},\\
\end{array}
\end{equation}
with a function $\nu(t)$ which depends on the system's
dynamics. The approximation is similar to coarse graining of
the classical distribution function. Correspondingly,
\begin{equation}\label{PoiW_0}
W_0(I;t)=\frac{2}{\hbar}\tanh\left(\frac{\nu}{2}\right)
\exp\left[-\frac{2}{\hbar}\tanh\left(\frac{\nu}{2}\right)\,I\right]\,.
\end{equation}

The dependence on time of the mean value of a function $Q({\hat
I})$ of the action operator ${\hat I}=\hbar {\hat n}$ is then
computed with the help of the formula
\begin{equation}\label{AverAcn}
\langle Q\rangle_t=\int_0^{\infty} dI Q(I) W_0(I;t)
\end{equation}
where the phase-space image of the operator ${\hat Q}$ is
calculated similarly to the zero harmonic amplitude $W_0(I;t)$
of the Wigner function (see Eq.~(\ref{Qfunc})). For example the
image of the operator ${\hat G}(\kappa)= e^{-\kappa{\hat I}}$,
which generates the moments of the amplitude $W_0(I;t)$ is
easily found to be
\begin{equation}\label{PhImG}
G(\kappa;I)=\frac{\,e^{\kappa\hbar/2}}{\cosh\left(\kappa\hbar/2\right)}\,
\exp\left[-2\tanh\left(\kappa\hbar/2\right)\frac{I}{\hbar}\right]\,.
\end{equation}
The mean value of the generating function of the action momenta
is readily obtained from eq. (\ref{AverAcn})
\begin{equation}\label{MeanG}
\langle G(\kappa;I)\rangle_t=e^{\frac{\kappa\hbar}{2}}\,
\frac{\sinh(\nu/2)}{\sinh(\nu/2+\kappa\hbar/2)}\,.
\end{equation}
In particular, the mean action equals $\langle
I\rangle_t=-\partial \langle
G(\kappa;I)\rangle_t/\partial\kappa\big|_{\kappa=0}=\frac{\hbar}{2}
\left[\coth\left(\frac{\nu}{2}\right)-1\right]$. This formula
relates the function $\nu(t)$ to the evolution of the action,
$e^{-\nu(t)}=\langle I\rangle_t/(\langle I\rangle_t+\hbar)$,
and allows us to represent finally the coarse-grained amplitude
of the zero harmonic in the form
\begin{equation}\label{PoiW_0+}
W_0(I;t)=\frac{1}{\langle I\rangle_t+\frac{\hbar}{2}}
\exp\left(-\frac{I}{\langle
I\rangle_t+\frac{\hbar}{2}}\right)\,.
\end{equation}
The time dependence of the mean action $\langle I\rangle_t$ is
shown in Fig.~\ref{fig:avn}

\subsection{Evolution of Nonzero Momenta: Complexity of Quantum States}

\label{sec:complexity}

The paramount property of classical dynamical chaos is the
exponentially fast structuring of the system's phase space on
finer and finer scales. In particular, the number ${\cal M}(t)$
of angular harmonics, that is the number of appreciably large
harmonic's amplitudes ${W^{(c)}}_{\,m}(I;t)$ in the Fourier
expansion (\ref{Four}) of the classical distribution function
$W^{(c)}(\alpha^*,\alpha;t)$ grows exponentially in time.
Namely ${\cal M}(t)\propto e^{t/\tau_c}$, where the
characteristic time $\tau_c$ goes to infinite when the
classical Lyapunov exponent vanishes. A simple consideration
shows that the exponential regime cannot last long in the case
of quantum dynamics. Indeed, in the terms of our auxiliary
two-dimensional linear oscillator where the functions $e^{i
m\theta}$ are the eigenstates of the operator ${\hat M}$
defined in (\ref{Moperator}) the mean number of harmonics
${\cal M}(t)\sim\langle|n_+ -n_-|\rangle_t \lesssim\langle
N\rangle_t=\langle n_+ +n_-\rangle_t \sim\langle
I\rangle_t/\hbar\,.$ Therefore the exponential upgrowth is
possible only for $e^{t/\tau_c}< \langle I\rangle_t/\hbar$,
namely for a time $t\lesssim t_E=\tau_c\ln\frac{\langle
I\rangle_t}{\hbar}\,.$ Since the mean action increases only
linearly in time, $t_E$ is basically the Ehrenfest
time~\cite{Berman78}, logarithmically short in $\hbar$.

In order to ascertain how complex the quantum state became by
the time $t$ we use as a probe a phase plane rotation by the
angle $\delta\theta\equiv\xi$. Such a rotation is generated by
the unitary transformation ${\hat P}(\xi)=e^{-i\xi{\hat n}}$ of
the density matrix ${\hat\rho(t)}$. The effect of such
perturbation, as mentioned above, is characterized by the Peres
fidelity (\ref{PFid}).

From the second line of Eq.~(\ref{PFid}) we obtain
\begin{equation}\label{FidRot}
F(\xi;t)=\sum_{n,n'=0}^{\infty}\cos[\xi(n'-n)]
\frac{\Big|\langle n'\big|{\hat\rho}(t)\big|n\rangle
\Big|^2}{\sum_{k=0}^{\infty}\langle k\big|{\hat\rho}^2(t)
\big|k\rangle}\,.
\end{equation}
The diagonal ($n'=n$) contribution
\begin{equation}\label{Fdiag0}
F_0(t)=\sum_{n=0}^{\infty} \frac{\Big|\langle
n\big|{\hat\rho}(t)\big|n\rangle\Big|^2}
{\sum_{k=0}^{\infty}\langle k\big|{\hat\rho}^2(t)\big|k\rangle}
\end{equation}
does not depend on $\xi$. Taking into account that
$F(\xi=0;t)=1$, we relate this contribution to the non-diagonal
part of the density operator:
\begin{equation}\label{Fdiag1}
F_0(t)=1-2\sum_{n'>n=0}^{\infty} \frac{\Big|\langle
n'\big|{\hat\rho}(t)\big|n\rangle\Big|^2}
{\sum_{k=0}^{\infty}\langle k\big|{\hat\rho}^2(t)\big|k\rangle}.
\end{equation}
After substitution of this expression into (\ref{FidRot}) we obtain
\begin{equation}\label{Fid_t}
F(\xi;t)=1-2\sum_{m=1}^{\infty}\sin^2\left(\xi m/2\right)
\mathcal{W}_m(t),
\end{equation}
where
\begin{equation}\label{mProb}
\mathcal{W}_m(t)=(2-\delta_{m0})\sum_{n=0}^{\infty}
\frac{\Big|\langle n+m\big|{\hat\rho}(t)\big|n\rangle\Big|^2}
{\sum_{n'=0}^{\infty}\langle
n'\big|{\hat\rho}^2(t)\big|n'\rangle}\,.
\end{equation}
Since
\begin{equation}\label{}
\begin{array}{c}
\sum_{n=0}^{\infty}\langle n\big|{\hat\rho}^2(t)\big|n\rangle=
\sum_{n=0}^{\infty}\langle n\big|{\hat\rho}(t)\big|n\rangle^2+\\
2\sum_{m=1}^{\infty}\sum_{n=0}^{\infty}\Big|\langle
n+m\big|{\hat\rho}(t)\big|n\rangle\Big|^2,
\end{array}
\end{equation}
we can convert Eq.~(\ref{mProb}) into
\begin{equation}\label{mProbConv}
\begin{array}{c}
{\mathcal{W}_{m}(t)}=\\
\frac{(2-\delta_{m0})\sum_{n=0}^{\infty}{\Big|\langle n+m\big|
{\hat\rho}(t)\big|n\rangle\Big|^2}\Big/
{\sum_{n'=0}^{\infty}\rho_{n' n'}^2(t)}}
{1+2\sum_{m=1}^{\infty}\sum_{n=0}^{\infty}\Big|\langle n+m\big|
{\hat\rho}(t)\big|n\rangle\Big|^2\Big/
{\sum_{n'=0}^{\infty}\rho_{n' n'}^2(t)}},
\end{array}
\end{equation}
where $\rho_{n n}(t)$ is a shorthand notation for $\langle n\big| {\hat\rho} (t)\big|n\rangle$.

Equivalently, one can express ${\mathcal{W}_{m}(t)}$ in terms
of the amplitudes (\ref{Harm}) of the Wigner function:
\begin{equation}\label{Harm_W}
\begin{array}{c}
\mathcal{W}_m(t)=\\
\frac{(2-\delta_{m0})\int_0^{\infty}dI\big|W_m(I;t)\big|^2
\big/\int_0^{\infty}dI
W_0^2(I;t)}{1+2\sum_{m=1}^{\infty}\int_0^{\infty}dI\big|W_m(I;t)\big|^2
\big/\int_0^{\infty}dI W_0^2(I;t)}\,.\\
\end{array}
\end{equation}
The completeness condition
\begin{equation}\label{Lcomp}
\begin{array}{c}
\sum_{n=0}^{\infty}\frac{(n+m)!}{n!}L_n^m(4I/\hbar)L_n^m(4I'/\hbar)=\\
\frac{\hbar}{4}\left(4I/\hbar\right)^{-m}e^{4I/\hbar}\delta(I-I')
\end{array}
\end{equation}
has been taken into account in deriving formula (\ref{Harm_W}).
Similarly to Eq.~(\ref{FidW}), expression (\ref{Harm_W})
remains valid in the classical limit, provided that the
harmonics of the classical distribution function are used.

\begin{figure}
\includegraphics[height=7.cm,angle=90]{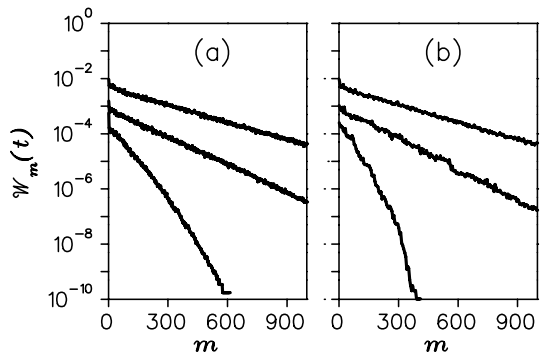}
\caption{Distribution of harmonics $\mathcal{W}_m (t)$ as a
function of $m$, with parameter values as in
Fig.~\ref{fig:rhodiag}.} \label{fig:Wmdistribution}
\end{figure}

Since the normalization condition $\sum_m\mathcal{W}_m(t)=1$
holds, the quantities $\mathcal{W}_m(t)$, $m\ge 0$ give the
probability distribution over the harmonic's numbers $m$. Now,
in the spirit of the linear response theory, we consider an
infinitesimally small rotation angle $\xi\rightarrow 0$ and
hold only the linear term of the power expansion of the density
operator. The Eq.~(\ref{Fid_t}) reduces then to
\begin{equation}\label{Aver_m}
\begin{array}{c}
F(\xi;t)\approx
1-\frac{1}{2}\xi^2\,\langle m^2\rangle_t\,,\\
\langle m^2\rangle_t=-\frac{d^2 F(\xi;t)}{d\xi^2}\Big|_{\xi=0}=
\sum_{m=1}^{\infty} m^2 \mathcal{W}_m(t)\,.
\end{array}
\end{equation}
Numerical simulations (see Fig.~\ref{fig:Wmdistribution}) show
that, apart from small fluctuations, the distribution
$\mathcal{W}_m(t)$ decreases with $m$ monotonically and
exponentially. Therefore the quantity $\sqrt{\langle
m^2\rangle}_t$ gives an estimate of the number ${\cal M}$ of
harmonics developed up to time $t$ and can be considered as a
suitable measure of complexity of the Wigner function at the
time $t$.

According to the first line of Eq.~({\ref{PFid}}), the same
fidelity can, alternatively, be presented in the form
\begin{equation}\label{FrevExp}
\begin{array}{c}
F(\xi,t)=\sum_{n,n'=0}^{\infty}\frac{\rho_n\rho_{n'}}
{\sum_{k=0}^{\infty}\rho_k^2}\Big|\langle n\big|e^{-i\xi{\hat
n}(t)}\big|n'\rangle\Big|^2= \\
1-\xi^2\,\left[\overline{\chi_2(t)} - 2\overline{\phi(t)}\right]
+O({\xi}^4)
\end{array}
\end{equation}
where the quantity
\begin{equation}\label{W_m_dis}
\overline{\chi_2(t)}=\sum_n\frac{\rho_n^2}{\sum_{n'}\rho_{n'}^2}
\left(\langle
n|{\hat n}^2(t)|n\rangle-\langle n|{\hat n}(t)|n\rangle^2\right)
\end{equation}
is the weighted-mean value of the standard deviation of the
excitation numbers at the moment $t$ whereas
$\overline{\phi(t)}$ is similar weighted-mean value of the
cumulative contribution
\begin{equation}\label{phi}
\phi(t)=\sum_{m=1}^{\infty}\left(1+\hbar/\Delta\right)^{-m}\,
\big|\langle n+m|{\hat n}(t)|n\rangle\big|^2
\end{equation}
of the off-diagonal matrix elements. Comparing now the
$\xi^2$-terms in the both equivalent representations
(\ref{Aver_m}) and (\ref{FrevExp}) of the fidelity we arrive at
the following {\it significant exact relation} between the time
behavior of the mean number of harmonics (which characterizes
the complexity) on the one hand and of the excitation numbers
on the other hand
\begin{equation}\label{m_ver_n}
\langle m^2\rangle_t=2\left[\overline{\chi_2(t)}-2\overline{\phi(t)}\right].
\end{equation}
The negative second contribution which appears only in the case
of mixed initial states reduces the number of harmonics.

\section{Quantum Reversibility: the Fidelity}

To explore the reversibility of the quantum dynamics we
consider now the perturbation angle $\xi$ as a free parameter
and investigate the differences between the initial and
reversed states as a function of $\xi$. To this end we first
analyze the fidelity $F(\xi;t)$ at an arbitrary moment $t$ as a
function of $\xi$.

Using Eq.~(\ref{Fid_t}), we can expand $F(\xi,t)$ in terms of
the even moments $\langle m^{2k} \rangle$ ($k=1,2,...$) of the
probability distribution (\ref{mProb})
\begin{equation}\label{FExpan}
F(\xi,t)=1-\sum_{k=1}^{\infty}(-1)^{k+1}
\frac{\xi^{2k}}{(2k)!}\langle m^{2k}\rangle_t.
\end{equation}
We remark that, due to the exponential decay (as a function of m) of the
distribution $\mathcal{W}_m(t)$, all moments $\langle
m^{2k}\rangle_t$ are finite.

On the other hand, the expansion of the general expression
(\ref{FrevExp}) over the parameter $\xi$ contains on- and
off-diagonal matrix elements of the powers of the operator
${\hat I}(t)=\hbar\,{\hat n}(t)$ and therefore connect the
Peres fidelity to the action evolution. Both the equivalent
representations (\ref{FExpan}) and (\ref{FrevExp}) will be
exploited below.

\vspace{1cm}

\subsection{Pure Coherent Initial State}

The theoretical analysis is especially easy to carry out in the
simple case of the (pure) ground initial state
$\rho_n=\delta_{n 0}$. The expression (\ref{FrevExp}) reduces
then to
\begin{equation}\label{PFidP}
F(\xi,t)=\big|\langle 0|{\hat
P(\xi,t)}|0\rangle\big|^2=
\Big|\langle 0|\,e^{-i\xi{\hat n}(t)}\,
|0\rangle\Big|^2\,.
\end{equation}
This specific initial state is the isotropic coherent state
$|\overset{\circ}\alpha=0\rangle$ which corresponds to the
distribution ${\cal
P}(|\overset{\circ}\alpha|^2)=\delta^{(2)}(\overset{\circ}\alpha)$
in Eq. (\ref{IsInMix}). After first few kicks, a state of
practically general form is produced.

\begin{figure}
\includegraphics[width=7.cm,angle=0]{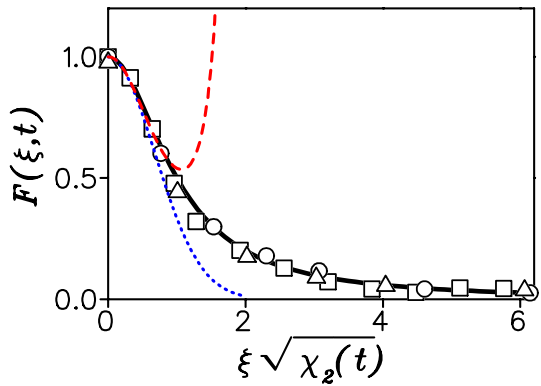}
\caption{(color online) Fidelity $F(\xi,t)$ versus perturbation
strength $\xi$, at $\hbar=0.25$, $g_0=2$ at three different
times: $t=5, \sqrt{\chi_2(t)}=80$ (squares); $t=25,
\sqrt{\chi_2(t)}=384$ (circles); $t=75, \sqrt{\chi_2(t)}=1010$
(triangles). The dotted and dashed curves show decay of the
r.h.s. in Eq. (\ref{FCumExp}) with, respectively, only the
lowest and the two lowest terms of the cumulant expansion being
kept. The full curve corresponds to the theoretical prediction
of the second line in Eq.~(\ref{FFidP}).} \label{fig:cumulants}
\end{figure}

Making use of the cumulant expansion we obtain
\begin{equation}\label{FCumExp}
F(\xi;t)=
\exp\left[-2\sum_{l=1}^{\infty}\frac{(-1)^{l-1}}{(2l)!}
\xi^{2l}\chi_{2l}(t)\right],
\end{equation}
where the cumulants (connected momenta) are
\begin{equation}\label{CumExpF}
\begin{array}{c}
\chi_2(t)=\langle 0|\left({\hat n}(t)-
\langle 0|{\hat n}(t)
|0\rangle\right)^2|0\rangle, \\
\chi_4(t)=\langle 0|\left({\hat n}(t)-
\langle 0|{\hat n}(t)
|0\rangle\right)^4|0\rangle-\\
3\left[\langle 0|\left({\hat n}(t)-
\langle 0|{\hat n}(t)|0
\rangle\right)^2|0\rangle
\right]^2,\\
\end{array}
\end{equation}
and so on. Correspondingly, $L(\xi;t)\equiv -\ln
F(\xi;t)=\xi^2\chi_2(t) -\frac{1}{12}\xi^4\chi_4(t)+...\,$. At
a given time $t$ we can retain only the lowest cumulant,
$L(\xi;t)\approx\xi^2\chi_2(t)$, as long as the perturbation
strength $\xi\lesssim 2\sqrt{3\chi_2(t)/\chi_4(t)}$. However,
as shown in Fig.~\ref{fig:cumulants}, this approximation fails
for larger values of $\xi$.

To go beyond such a restricted range of values of the parameter
$\xi$ we observe that the amplitude $f(\xi;t)=\langle
0|e^{-i\xi{\hat n}(t)}|0\rangle$ can be readily represented as
\begin{equation}\label{FidAmp}
f(\xi;t)=\sum_{n=0}^{\infty}w_n(t)\,e^{-i\xi n},
\end{equation}
where $w_n(t)\equiv\langle n|{\hat
\rho}(t)|n\rangle=\big|\langle n|{\hat U} (t) |0\big|^2$ is the
excitation number probability distribution.

The probability $w_n(t)$  exhibits larger fluctuations as a
function of $n$ than those in the case of broad initial
mixtures (compare the two panels of Fig.~\ref{fig:rhodiag}
obtained for pure (right) and mixed (left) initial
distributions). However,  at any given time $t$ larger than the
Ehrenfest time $w_n(t)$ decays, on average, exponentially.
Assuming, similarly to (\ref{DiagAnz}), the exponential ansatz
$w_n(t)\approx\left[1-e^{-\lambda(t)}\right]\,e^{-\lambda(t)n}$,
we obtain
\begin{equation}\label{DiagAnzP}
w_n(t)\approx\frac{1}{\langle n\rangle_t+1}\, \left[\frac{\langle
n\rangle_t}{\langle n\rangle_t+1}\right]^n,
\end{equation}
where $\langle n\rangle_t$ is the mean excitation number at the
moment $t$. The approximation (\ref{DiagAnzP}) leads to the
result
\begin{equation}\label{FFidP}
\begin{array}{c}
F(\xi;t)\approx\frac{1}{1+4\langle n\rangle_t \left(\langle
n\rangle_t+1\right)\,\sin^2(\xi/2)}=\\
\frac{1}{1+4\chi_2(t)\,\sin^2(\xi/2)}\approx
\frac{1}{1+\xi^2\chi_2(t)}\,.
\end{array}
\end{equation}
In the second line we have taken into account that
approximation (\ref{DiagAnzP}) implies the following relation:
\begin{equation}\label{ReltoCum}
\chi_2(t)\equiv\langle 0|n^2|0\rangle_t -\langle
0|n|0\rangle_t^2 \approx\langle n\rangle_t \left(\langle
n\rangle_t+1\right).
\end{equation}
(Let us remind in this connection that we have chosen
$\overset{\circ}\alpha =0$). Opposite to the exact relation
$\chi_2(t)=\frac{1}{2}\langle m^2\rangle_t$ (see eq.
(\ref{m_ver_n})) the relation eq. (\ref{ReltoCum}) is valid
only after the Ehrenfest  time $t_E$. Notice that accordingly
to the first line in Eq.~(\ref{FFidP}) $F(2\pi;t)=F(0;t)=1$, as
expected since rotation by the angle $\xi=2\pi$ is just the
identity operation.

Fig.~\ref{fig:cumulants} shows that the fidelity decay is
nicely described by the analytical formula (\ref{FFidP}). It is
also clearly seen that the fidelity almost vanishes already at
very small values of the parameter $\xi$, so that the
approximation given in the second line of Eq.~(\ref{FFidP})
works quite well. We can therefore conclude that in the
considered case of the pure ground initial state only the
lowest cumulant $\chi_2$ determines the overall decay of the
fidelity, whereas the higher ones are responsible for the
fluctuations in the decay law.

Using the above exponential ansatz, the expression
(\ref{mProb}) for the probability distribution
$\mathcal{W}_m(t)$ simplifies to
\begin{equation}\label{mProbP}
\begin{array}{c}
\mathcal{W}_m(t)=(2-\delta_{m 0})
\sum_{n=0}^{\infty}w_{n+m}(t)\,w_n(t)\\
\approx\frac{2-\delta_{m0}}{2\langle
n\rangle_t+1}\,\left[\frac{\langle n\rangle_t}{\langle
n\rangle_t+1}\right]^m,
\end{array}
\end{equation}
so that this distribution decays with the same slope
$\lambda(t)$ as the distribution of the excitation numbers
(\ref{DiagAnzP}). Together with Eq.~(\ref{ReltoCum}) this
yields the following relation:
\begin{equation}\label{PAver_m}
\langle m^2\rangle_t=2\,\chi_2(t)\approx 2\langle
n\rangle_t\,(\langle n\rangle_t+1)\,.
\end{equation}
Therefore, after an initial interval of order of the Ehrenfest
time, the number of harmonics of the Wigner function increases
with time in the same manner as the excitation number, not
faster than linearly. Notice also that, as it should be,
substitution of the expression (\ref{mProbP}) in the general
formula (\ref{Fid_t}) leads again to the same result
(\ref{FFidP}).

\begin{figure}
\includegraphics[width=7.cm,angle=0]{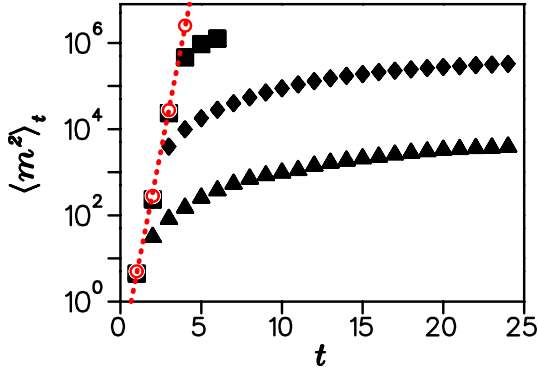}
\caption{(color online) $\langle m^2 \rangle_t$ versus $t$, at
$g_0=1.5$, $\delta=0.5$. Squares, diamonds and triangles
correspond to $\hbar = 0.01, 0.1$ and 1. In this latter case,
the initial condition corresponds to the ground state,
$\hat{\rho}(0)=|0\rangle\langle 0|$. Empty circles refer to
classical dynamics and the dashed line is an exponential fit to
these data, $\langle m^2\rangle_t=\exp(a+b t)$, with
$a\approx-2.9$ and $b\approx4.4$.} \label{fig:Wquantclass}
\end{figure}

The relatively slow dependence of $F(\xi;t)$ and the number of
harmonics ${\cal M}(t)=\sqrt{\langle m^2\rangle_t}$ on time,
which follows from expressions (\ref{FFidP}) and
(\ref{PAver_m}) should be juxtaposed with the classical
behavior dictated by the exponential instability of the
classical dynamics. The latter manifests itself, in particular,
in the exponential growth of the number of harmonics of the
classical phase-space distribution function
$W_c(\alpha^*,\alpha;t)$. To accomplish such a comparison we
solve the classical Liouville equation with the initial phase
space distribution $W_c(\alpha^*,\alpha;0)=\frac{1}{\delta}\,
e^{-\frac{|\alpha|^2}{\delta}}$ of size $\delta$ which
coincides, for a given value of $\hbar$, with the size
$\hbar/2$ of the Wigner function corresponding to the initial
quantum ground state ${\hat{\rho}}(0)=|0\rangle\langle 0|$. The
quantum to classical transition is explored by keeping $\delta$
constant and considering, for smaller and smaller values of
$\hbar$, initial incoherent mixtures (\ref{InWfunc}) of size
$\delta=\Delta+\hbar/2$. A numerical illustration of such a
procedure is presented in Fig.~\ref{fig:Wquantclass}. The
exponential increase of $\langle m^2 \rangle_t$ up to the
Ehrenfest time is clearly seen. After that time, a much slower
power-law increase follows, in accordance, for pure states,with
the relation (\ref{PAver_m}). Such a behavior is consistent
with the findings reported in the Refs.~\cite{gu,brumer}.

\begin{figure}
\includegraphics[height=7.cm,angle=90]{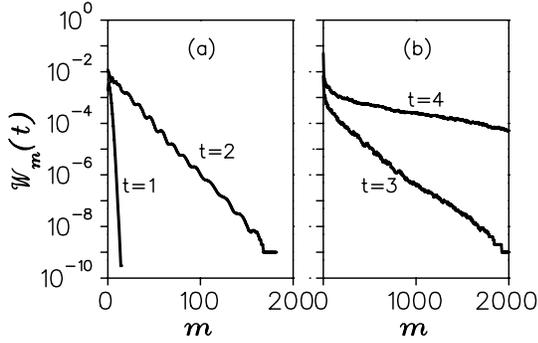}
\caption{Distribution $\mathcal{W}_m$ of harmonics at times
smaller than the Ehrenfest time, for parameter values
$\hbar=0.01$, $g_0=1.5$. Data at $t=1$ and $t=3$ are scaled by
a factor $0.1$.} \label{fig:Ehrenfest}
\end{figure}

The first relation in Eq.~(\ref{PAver_m}) allows us to directly
connect the Peres fidelity after the Ehrenfest time with the,
characterized by the mean number $\sqrt{\langle m^2\rangle_t}$
of harmonics, complexity of the Wigner function at time $t$:
\begin{equation}\label{FidPVerCom}
\begin{array}{c}
F(\xi;t)\approx\frac{1}{1+\frac{1}{2}\xi^2\langle m^2\rangle_t} =\\
\,\,\,\,\,\,\,\\
1-\sum_{k=1}^{\infty}(-1)^{k+1}\xi^{2k}(\langle
m^2\rangle_t)^k\,.
\end{array}
\end{equation}
More than that, comparing the terms of the two expansions,
(\ref{FExpan}) and (\ref{FidPVerCom}), the former of which is
exact for any time including times shorter or of the order of
the Ehrenfest time, we see that also the latter would be always
correct if
\begin{equation}\label{PoisRel}
\langle m^{2k}\rangle_t=\frac{(2k)!}{2^k}(\langle
m^2\rangle_t)^k\,.
\end{equation}
Such a relation is characteristic of the exponentially decaying
distribution. Numerical data presented in
Fig.~\ref{fig:Ehrenfest} support such a conjecture. It follows
then that in accordance with the different growth of the number
of harmonics before and after the Ehrenfest time (see
Fig.~\ref{fig:Wquantclass}), the slope of the $m$-dependence of
the distribution $\mathcal{W}_m(t)$ is drastically different
inside and outside the Ehrenfest time scale. This slope
decreases exponentially with $t$ in the first case and not
faster than linearly in the second.

\section{Quantum Reversibility: the Time-Reversed State.}

In this section, we study in detail the phase-space structure
of the time-reversed state characterized by the harmonics
content of the Wigner function in dependence on the
perturbation strength $\xi$ and the reversal time $T$.

Generally, the Peres fidelity $F(\xi;T)$, as it appeares in the
first line of Eq.~(\ref{FidW}), is sensitive only to the
distortion of the zero harmonic of the reversed Wigner function
or, equivalently, to the redistribution of the excitation
numbers. Utilizing the first line of Eq.~(\ref{PFid}) we obtain
in the terms of the density matrix
\begin{equation}\label{MixedFid}
\begin{array}{c}
F(\xi;T)=\sum_{n=0}^{\infty}\frac{\rho_n}{\sum_{n'=0}^{\infty}\rho_{n'}^2}
\langle n\big|{\hat{\tilde\rho}}(0|T,\xi)\big|n\rangle\approx\\
2\sum_{n=0}^{\infty} e^{-\frac{\hbar}{\Delta}\,n}\, \langle
n\big|{\hat{\tilde\rho}}(0|T,\xi)\big|n\rangle,\,\,\,(\Delta\gg\hbar),
\end{array}
\end{equation}
where only the diagonal matrix elements
\begin{equation}\label{RevDiagP}
\begin{array}{c}
\langle n|{\hat{\tilde\rho}}(0|T,\xi)|n\rangle=
\sum_{n'=0}^{\infty}\rho_{n'}\,\Big|\langle n\big|{\hat P}(\xi,T)
\big|n'\rangle\Big|^2=\\
\rho_n\,\Big|\langle n\big|{\hat P}(\xi,T) \big|n\rangle\Big|^2+\\
\sum_{n'=0}^{\infty}\left(1-\delta_{nn'}\right)\rho_{n'}\,\Big|
\langle n\big|\left({\hat P}(\xi,T)-1\right)\big|n'\rangle\Big|^2
\end{array}
\end{equation}
of the reversed density matrix are present. Only the term which
stays in the second line of this equation remains in the limit
$\xi\rightarrow 0$. This limit is shown with the dotted line in
the Fig. 8.

The effect of perturbation shows up first in the second order
with respect to the perturbation parameter $\xi$. Expanding in
Eq.~(\ref{RevDiagP}) the operator ${\hat
P}(\xi,T)=e^{-i\,\xi{\hat n}(T)}$ up to the correction of this
order we find
\begin{equation}\label{SmallxiDiagME}
\begin{array}{c}
\langle n|{\hat{\tilde\rho}(0|T,\xi)}|n
\rangle\approx\\
\rho_n\left[1-\xi^2\left(\langle n|{\hat
n}^2(T)|n\rangle-\langle n|{\hat
n}(T)|n\rangle^2\right)\right]+\\
\xi^2\sum_{n'\neq n}\rho_{n'} \big|\langle n'|
{\hat n}(T)|n\rangle\big|^2
\end{array}
\end{equation}
which leads to Eq. (\ref{FrevExp}) again.

The fidelity (\ref{MixedFid}) consists of two different contributions
$\overline{F_n(\xi;T)}^{(n)}+\overline{F^{(m)}(\xi;T)}$ one of which
\begin{equation}\label{MeanFid}
\overline{F_n(\xi;T)}^{(n)}\equiv \sum_{n=0}^{\infty}\frac{\rho_n^2}
{\sum_{k=0}^{\infty}\rho_k^2}
\Big|\langle n\big|{\hat P}(\xi,T) \big|n\rangle\Big|^2
\end{equation}
is the weighted mean of pure state fidelities, whereas the
off-diagonal contribution
\begin{equation}\label{ExtraTermFid}
\begin{array}{c}
\overline{F^{(m)}(\xi;T)}=\\
2\sum_{n=0}^{\infty} \frac{\rho_n^2}
{\sum_{k=0}^{\infty}\rho_k^2}\sum_{m=1}^{\infty}
e^{-\frac{\hbar}{\Delta} m} \Big|\langle n+m\big|{\hat
P}(\xi,T)\big|n\rangle\Big|^2
\end{array}
\end{equation}
is specific for mixed initial states.

However, the important information on the harmonics that
survived the backward evolution is absent in the fidelity
(\ref{MixedFid}). To find their number and the corresponding
distribution $\tilde{\mathcal{W}}_{m}(0|\,\xi;T)$ we perturb
the reversed density matrix ${\hat{\tilde\rho}}(0|T,\xi)$ by
means of the probing operation ${\hat P}(\xi')=e^{-i\xi'{\hat
n}}$, with a new infinitesimally small rotation angle $\xi'$ .
In the same manner as in Sec.~\ref{sec:complexity} we obtain
(compare with (\ref{Aver_m}))
\begin{equation}\label{RevAver_m}
\langle {\tilde m}^2(\xi;T)\rangle= \sum_{m=1}^{\infty}m^2
\tilde{\mathcal{W}}_m(0|\,\xi;T),
\end{equation}
where
\begin{equation}\label{Rev_mProb}
\begin{array}{c}
\tilde{\mathcal{W}}_{m}(0|\,\xi;T)=
(2-\delta_{m0})\sum_{n=0}^{\infty}\frac{\Big|\langle n+m\big|
{\hat{\tilde\rho}}(0)\big|n\rangle\Big|^2}
{\sum_{k=0}^{\infty}\langle k\big|{\hat{\tilde\rho}}^2(0)
\big|k\rangle}=\\
\frac{(2-\delta_{m0})\sum_{n=0}^{\infty}{\Big|\langle n+m\big|
{\hat{\tilde\rho}}(0)\big|n\rangle\Big|^2}\Big/
{\sum_{k=0}^{\infty}{\tilde\rho}_{k
k}^2(0)}}{1+2\sum_{m=1}^{\infty}\sum_{n=0}^{\infty}\Big|\langle
n+m\big| {\tilde\rho}(0)\big|n\rangle\Big|^2\Big/
{\sum_{k=0}^{\infty}{\tilde\rho}_{k k}^2(0)}}=\\
\frac{(2-\delta_{m0})\int_0^{\infty}\!dI
\big|\tilde{W}_m(I;0)\big|^2\Big/\int_0^{\infty}\!dI\,
\tilde{W}_0^2(I;0)}{1+2\sum_{m=1}^{\infty}\int_0^{\infty}\!dI
\big|\tilde{W}_m(I;0)\big|^2\Big/\int_0^{\infty}\!dI\,
\tilde{W}_0^2(I;0)}\,.
\end{array}
\end{equation}
We have used the shorthands ${\hat{\tilde\rho}(0|T,\xi)}\Rightarrow
{\hat{\tilde\rho}(0)}$ and $\tilde{W}_m(I;0|T,\xi)\Rightarrow
\tilde{W}_m(I;0)$ in these formulae.\\

\subsection{Pure Initial State}

\begin{figure}
\includegraphics[width=7.cm,angle=0]{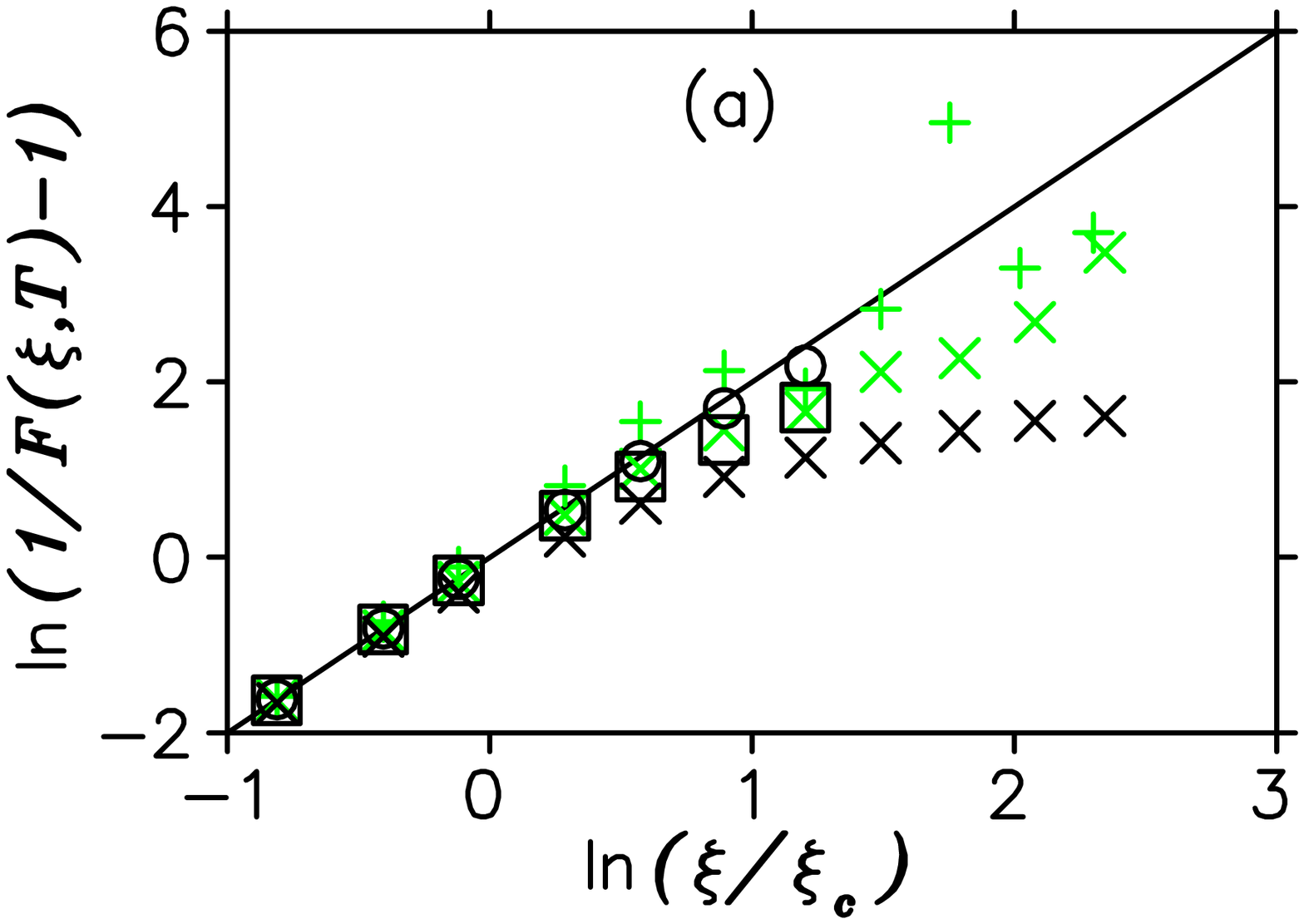}
\includegraphics[width=7.cm,angle=0]{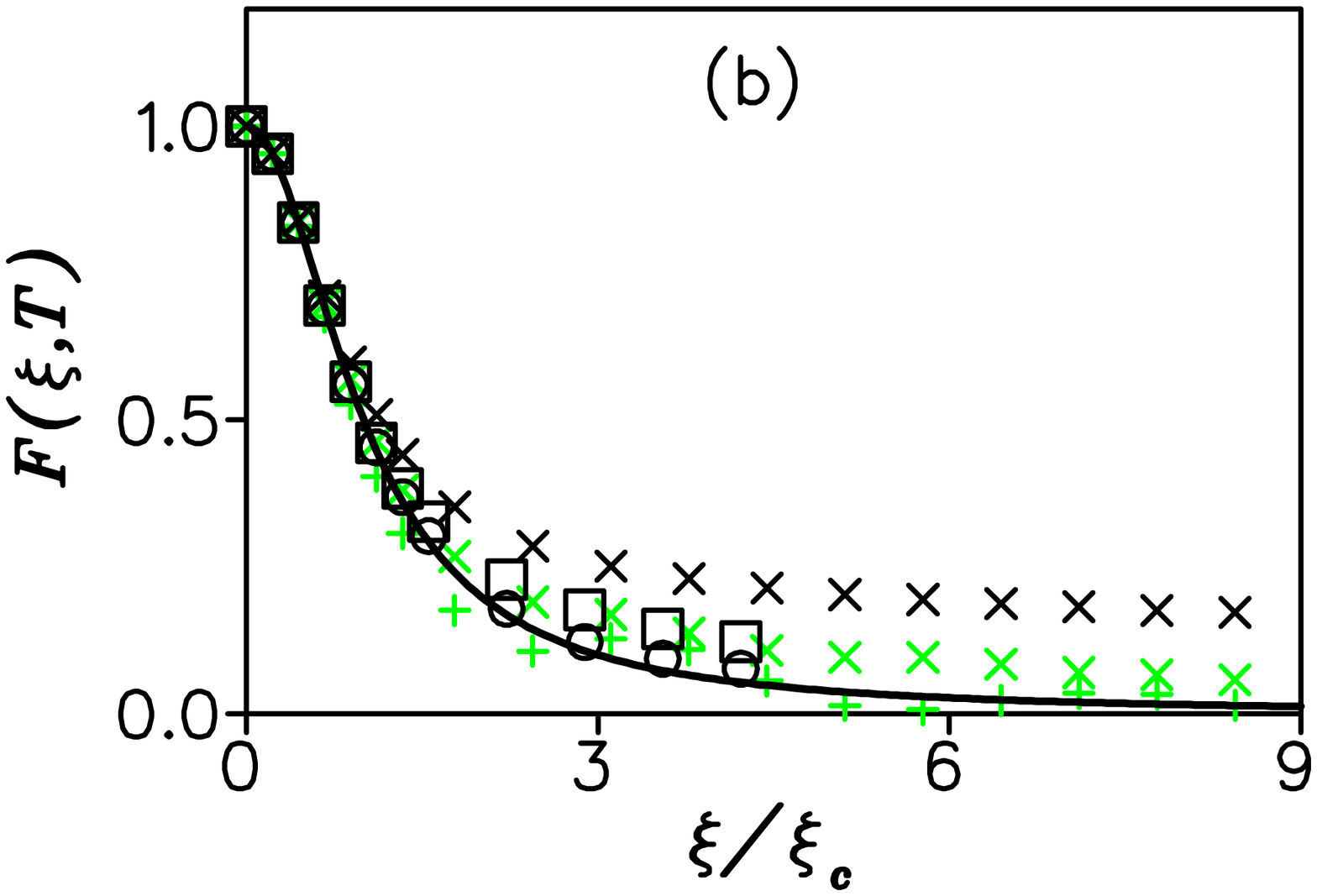}
\caption{(color online) Fidelity $F(\xi;T)$ versus the scaled
variable $\xi/{\xi}_c(T)$. Data correspond to: (i) $\hbar = 1,
g_0 = 2$; light pluses: $T = 10$, $\Delta = 0$; light and black
crosses: $T = 50$, $\Delta = 0$ and 25; (ii) $\hbar = 0.1, g_0
=1.5$, $\Delta = 0.45$ and $T=20$; (black open circles); (iii)
$\hbar = 0.01, g_0 = 1.5, \Delta =0.5$ and $T = 2$. (black open
squares). The curves show the theoretical prediction of
Eq.~(\ref{FidPVerCom}).} \label{fig:Fidvsn2}
\end{figure}

We note now that, according to the first line of the
relation~(\ref{FidPVerCom}), in the special case of a pure
initial state the crossover
\begin{equation}\label{Crssover}
F(\xi,T)\approx\left\{
\begin{array}{c}
1,\qquad \mbox{if} \quad \xi\ll\xi_c(T),\\
\frac{2}{\xi^2\,\langle m^2\rangle_T}, \qquad \mbox{if} \quad
\xi\gg\xi_c(T)
\end{array}\right.
\end{equation}
from good to poor reversibility takes place near the {\it
critical value} $\xi_c(T)\equiv\sqrt{2/\langle m^2\rangle_T}$
of the strength $\xi$ of the perturbation. The validity of the
formula (\ref{FidPVerCom}) for pure initial states is
illustrated by the numerical data plotted in
Fig.~\ref{fig:Fidvsn2}.

The numerical results presented in Fig.~\ref{fig:Wquantclass}
imply that the fidelity $F(\xi,T)$ decays as a function of the
reversal time $T$ the faster (approaching the exponential decay
typical of the classical chaotic dynamics \cite{benenti02}) the
closer the motion is to the semi-classical region. Within the
interval $T\lesssim$ the Ehrenfest time the decay is
exponential with the rate $1/\tau_c$ which describes the
classical exponential proliferation of the number of harmonics
and {\it does not depend on the perturbation constant $\xi$}.

With regard to the number of harmonics of the time-reversed
state, expression (\ref{Rev_mProb}) reduces, for a pure initial
state, (compare with Eq. (\ref{mProbP})) to
\begin{equation}\label{tildemProbP}
\tilde{\mathcal{W}}_{m}(0|\,\xi;T)=
(2-\delta_{m 0})
\sum_{n=0}^{\infty}{\tilde w}_{n+m}(\xi;T)\,{\tilde w}_n(\xi;T),\\
\end{equation}
where
\begin{equation}\label{tildeDiagRho}
{\tilde w}_n(\xi;T)\equiv
\tilde{\rho}_{n n}(0|T,\xi)=
\big|\langle n|e^{-i\xi{\hat n}(T)}|0
\rangle\big|^2\,.
\end{equation}
In particular ${\tilde w}_0(\xi;T)=F(\xi;T)$ (see Eq. (\ref{PFidP})).

\begin{figure}
\includegraphics[width=6.cm]{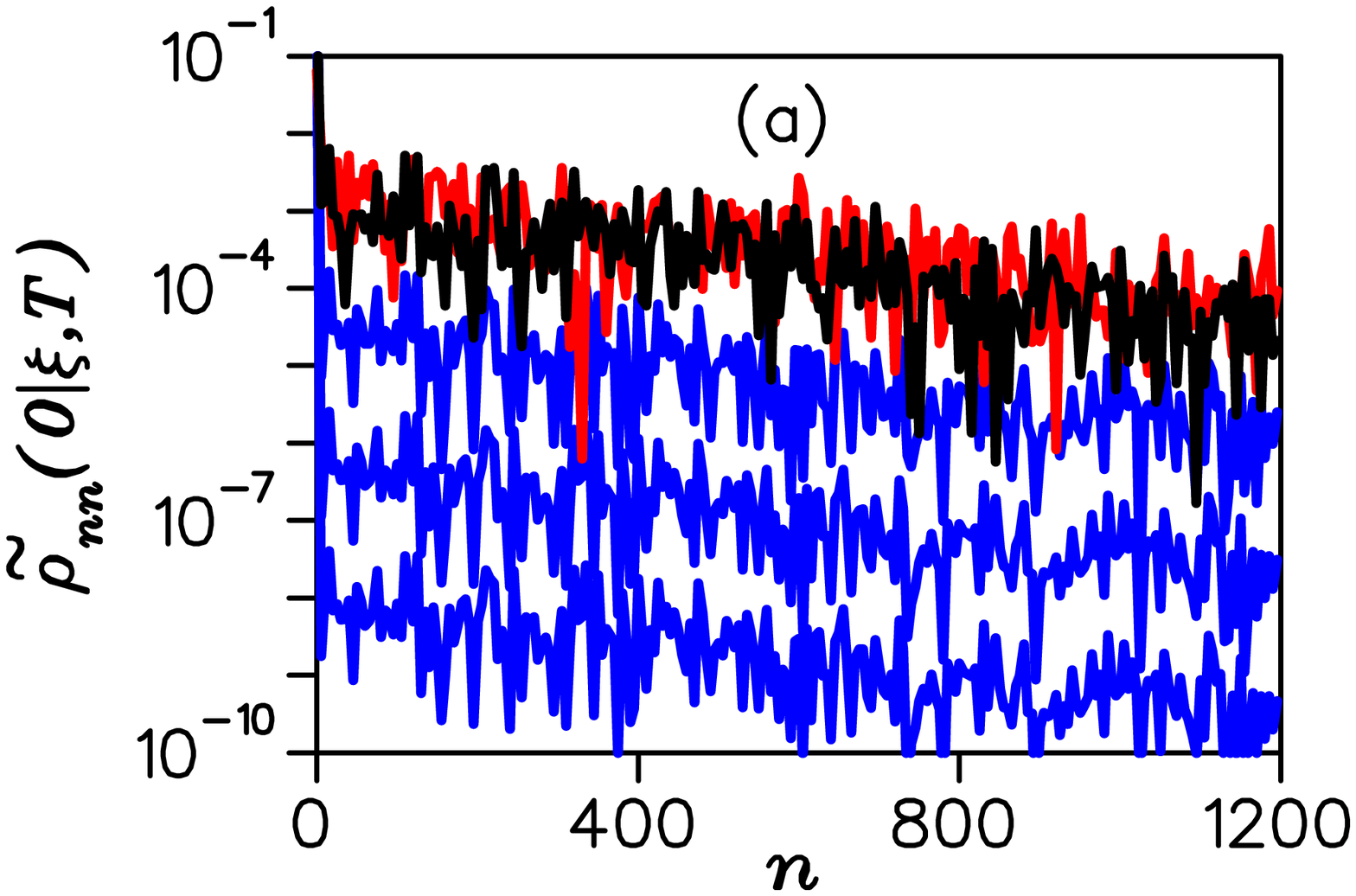}
\includegraphics[width=6.cm]{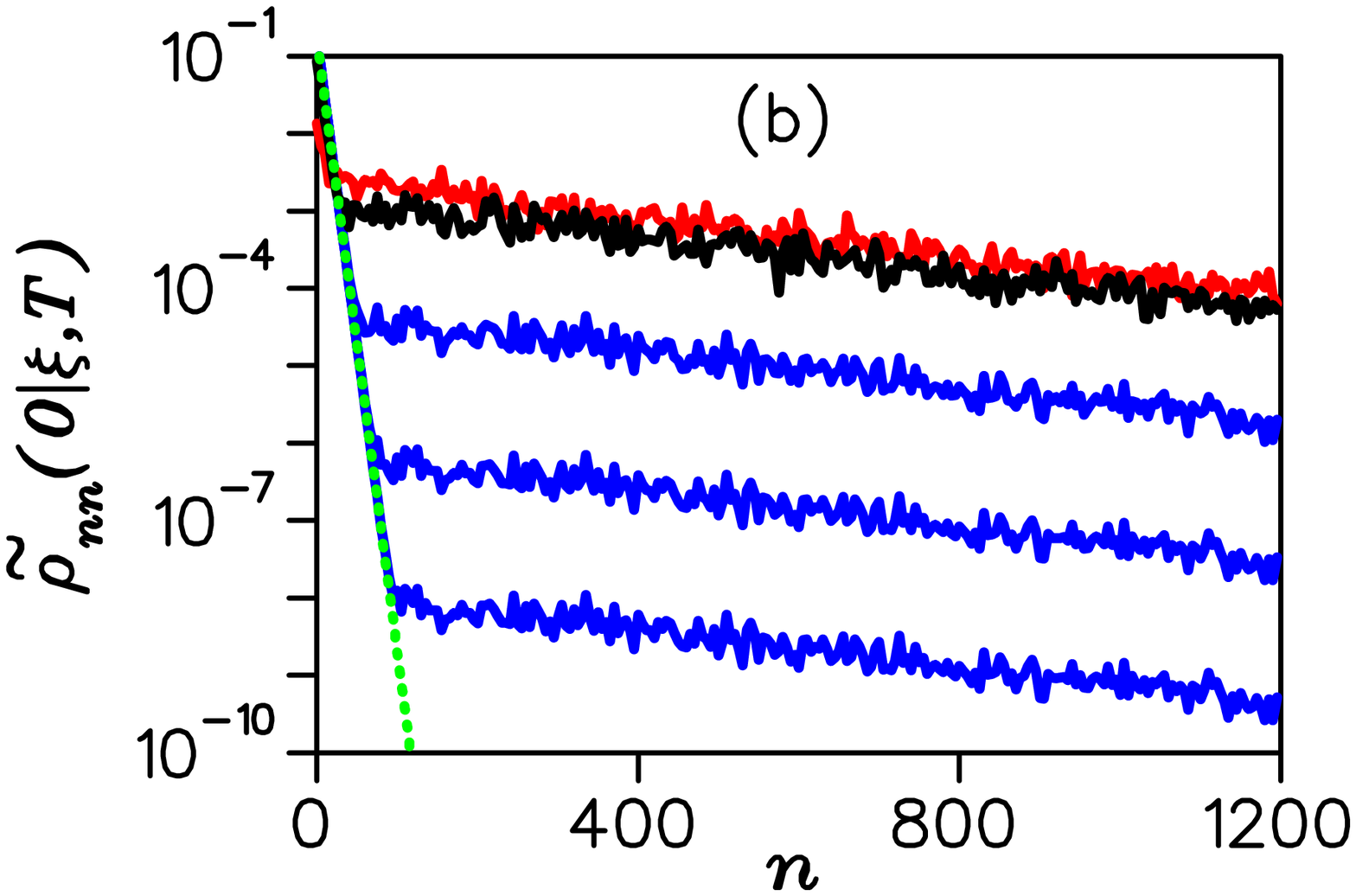}
\caption{(color online) Decay of $\tilde{\rho}_{n n}(0|\xi,T)$
as a function of $n$, at $\hbar=1$, $g_0=2$ and $T=50$, for (a) \textit{pure} ($\Delta=0$)
and (b) \textit{mixed} ($\Delta=5$) initial states. Curves from bottom to top
correspond to different values of perturbation parameter
$\xi=\xi_c\times 2^{l/2}$, $l=-9,-6,-3,0$ and $3$. The dotted
line corresponds to the initial distribution $\rho_{n
n}(0)=\frac{\hbar}{\Delta+\hbar}\,
\left(\frac{\Delta}{\Delta+\hbar}\right)^n$.}
\label{fig:expansatz}
\end{figure}

\begin{figure}
\includegraphics[width=6.cm]{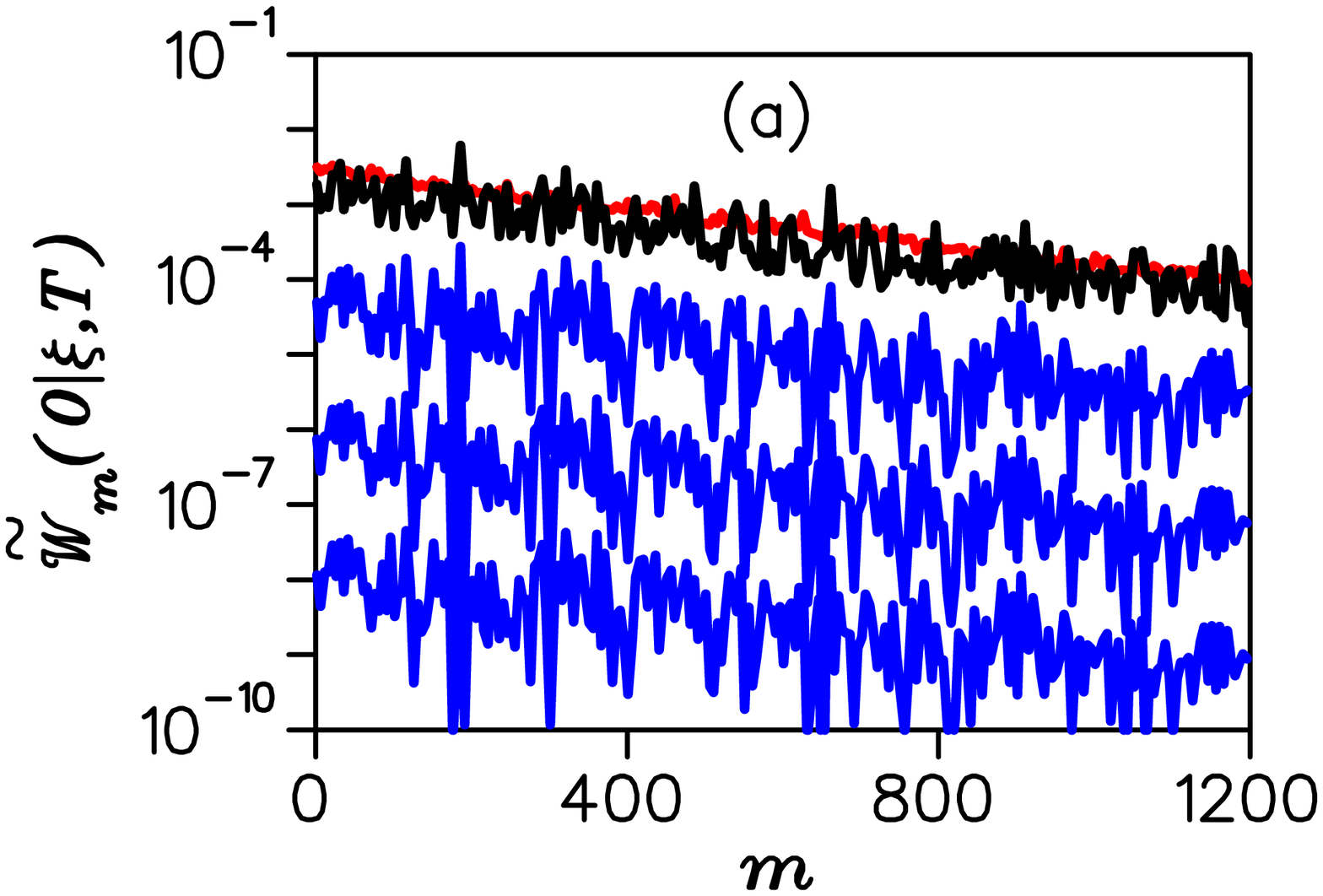}
\includegraphics[width=6.cm]{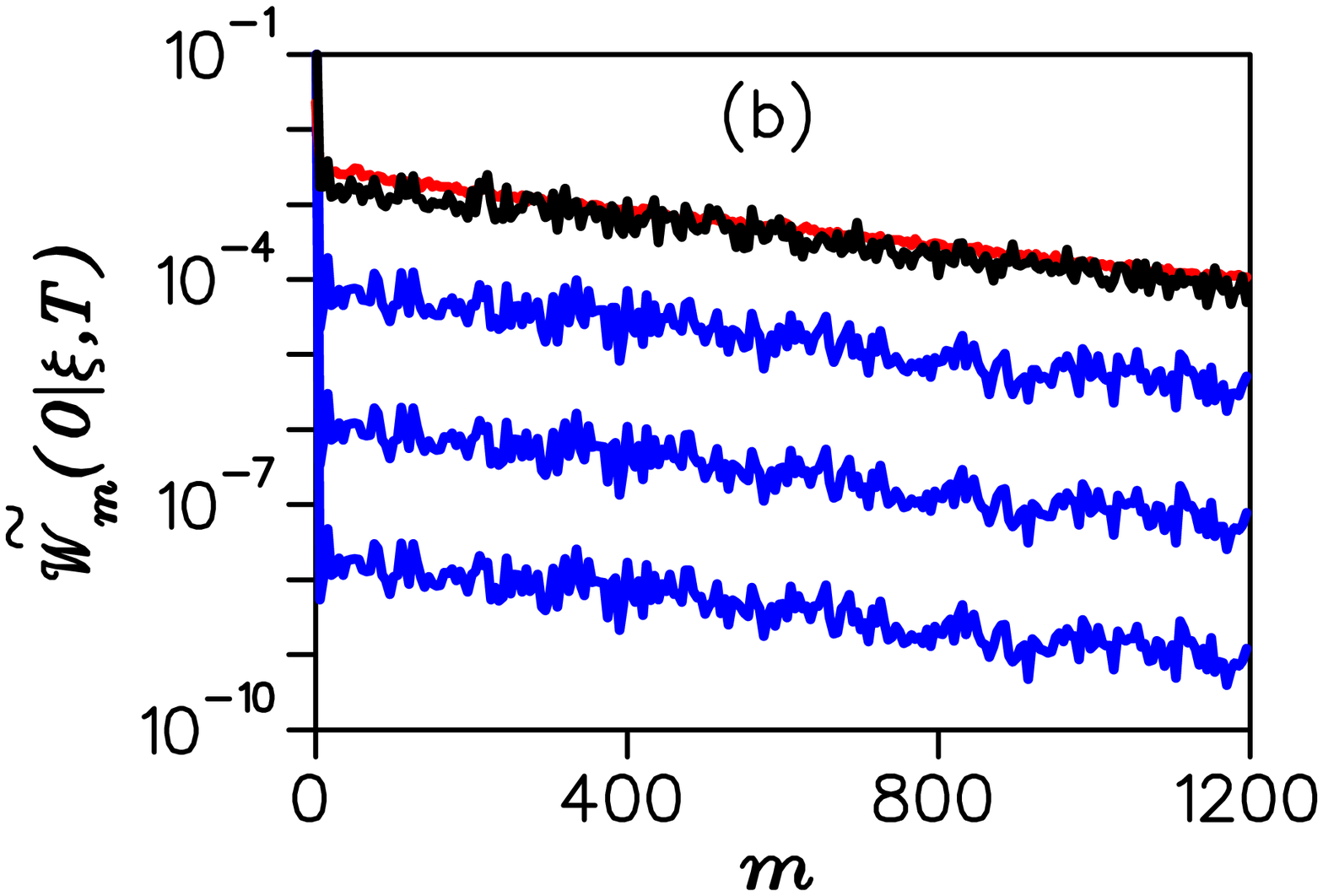}
\caption{(color online) Decay of ${\tilde{\cal W}}_m(0|\xi,T)$
as a function of $m$, with $\Delta=0$ and the other parameter
values as in Fig.~\ref{fig:expansatz}.} \label{fig:expansatz2}
\end{figure}

We expect again an overall exponential decay of the excitation
number probability distribution: ${\tilde w}_{n\geq
1}(\xi;T)\approx A(\xi;T)\,e^{-{\tilde\nu(T)}n}$. This
assumption is well confirmed by numerical simulations, examples
of which are presented in Figs.~\ref{fig:expansatz},
\ref{fig:expansatz2} where exponential decays of both the
$n$-and $m$-distributions ${\tilde w}_{n\geq 1}(\xi;T),\,\,
\tilde{\mathcal{W}}_{m}(0|\,\xi;T)$ is demonstrated in
agreement with Eq.~(\ref{tildemProbP}). It is clearly seen that
the decay rate $\tilde\nu(T)$ {\it does not depend on the
perturbation strength} $\xi$. The normalization condition
$\sum_{n=0}^{\infty}{\tilde w}_n=1$ defines the normalization
constant $A(\xi;T)$ thus connecting the $n$-distribution with
the fidelity
\begin{equation}\label{tilde_nProb}
{\tilde w}_{n\geq 1}(\xi;T)=\left(1-F(\xi;T)\right)\,
\left(e^{{\tilde \nu(T)}}-1\right)\,e^{{-\tilde \nu(T)}n}\,.
\end{equation}\\

A simple although a bit lengthy calculation connects, quite
similarly to the relation expressed by Eqs. (\ref{ReltoCum})
and (\ref{PAver_m}), the second moment of the distribution
(\ref{tildemProbP}) to the fluctuations of the excitation
numbers:
\begin{equation}\label{ResPAver_m}
\begin{array}{c}
\langle{\tilde m^2}(\xi;T)\rangle=
2\left[\langle{\tilde n^2}(\xi;T)\rangle-
\langle{\tilde n}(\xi;T)\rangle^2\right]=\\
2\frac{\left(1-F(\xi;T)\right)\left(F(\xi;T)+e^{-{\tilde \nu(T)}}\right)}
{\left(1-e^{-{\tilde\nu(T)}}\right)^2}\,.\\
\end{array}
\end{equation}
Notice that the ratio of the first two moments of the
distribution (\ref{tildemProbP}) calculated with the help of
the parametrization (\ref{tilde_nProb})
\begin{equation}\label{Lambda}
\Lambda(T)\equiv\frac{\langle{\tilde
m}(\xi;T)\rangle}{\langle{\tilde m^2}
(\xi;T)\rangle}=\frac{\langle{\tilde
n}(\xi;T)\rangle}{\langle{\tilde
n^2}(\xi;T)\rangle}=\frac{1-e^{{-\tilde \nu(T)}}}{1+e^{{-\tilde
\nu(T)}}}
\end{equation}
does not depend on $\xi$ and is small if the reversal time $T$
is not too small. In this case $\tilde\nu(T)\approx
2\Lambda(T)\ll 1$.

When the perturbation parameter $\xi\ll\xi_c(T)$ so that the
Peres fidelity (\ref{FidPVerCom}) is close to 1, we obtain from
Eq. (\ref{ResPAver_m}) the ratio
\begin{equation}\label{m-s_Ratio}
\begin{array}{c}
R(\xi,T)\equiv\frac{\langle{\tilde
m^2}(\xi;T)\rangle}{\langle m^2\rangle_T}\approx
\frac{1}{2}\left(\frac{\xi}{\Lambda(T)}\right)^2\\
=\frac{1}{2}\left(\frac{\xi_c(T)}{\Lambda(T)}\right)^2\,
\left(\frac{\xi}{\xi_c(T)}\right)^2,
\end{array}
\end{equation}
which characterizes the residual complexity of the reversed
state. This ratio remains small and therefore the motion
remains practically reversible as long as $\xi\ll\Lambda(T)$.
For the parameters fixed in the top panel of Fig. 8 the ratio
$\xi_c(T)/{\sqrt{2}\Lambda(T)}\approx 1.44$ so that the
condition of reversibility looks as $\xi<0.69\,\xi_c(T)$ in
reasonable agreement with the condition obtained above with the
help of fidelity.

\subsection{Incoherent Initial Mixture}

When the evolution starts with a mixed initial state the
$\xi^2$-correction in the expansion (\ref{FrevExp}) of the
Peres fidelity contains along with the negative contribution
defined by the weighted-mean value of the lowest cumulant
$\overline{\chi_2(T)}$ (compare with Eq.~(\ref{ReltoCum})) an
additional positive one. Still, the condition
$\xi\lesssim{\xi}_c(T)$ holds as the criterion for the Peres
fidelity to be close to one.

Contrary to the diagonal matrix elements (\ref{SmallxiDiagME}),
the off-diagonal elements of the density matrix are distorted
already in the first order approximation,
\begin{equation}\label{RevRhoLinME}
\begin{array}{c}
\langle n+m|{\hat{\tilde\rho}(0|T,\xi)}|n\rangle\approx\\
\rho_n\delta_{m0}+i\xi\langle n+m|{\hat
n}(T)|n\rangle\,\rho_n\left(1-\frac{\rho_{n+m}}{\rho_n}\right)\,.
\end{array}
\end{equation}
As long as $\xi$ is small enough only the probability of zero
harmonic $\tilde{\mathcal{W}_0}$ is close to unity whereas the
probabilities of other harmonics are small as $\xi^2$:
\begin{equation}\label{Rev_mProbLin}
\begin{array}{c}
\tilde{\mathcal{W}}_{m\geqslant 1}(0|\,\xi;T)\approx
\xi^2\,{\tilde q}_m(0|0,T),\\
\tilde{\mathcal{W}}_0(0|\,\xi;T)\approx1-\xi^2\,{\tilde q}_0(0|0,T)\,,
\end{array}
\end{equation}
where
\begin{equation}\label{tildew_m}
\begin{array}{c}
{\tilde q}_{m\geqslant 1}(0|0,T)=
\left[1-\left(\frac{\Delta}{\Delta+\hbar}\right)^m\right]^2\times\\
\Sigma_{n=0}^{\infty}\frac{\rho_n^2}{\sum_{k=0}^{\infty}\rho_k^2}
|\langle n+m|{\hat n}(T)|n\rangle|^2\equiv d_m\times{\tilde
r}_m(0|0,T)
\end{array}
\end{equation}
and
\begin{equation}\label{tildew_0}
{\tilde q}_0(0|0,T)=\sum_{m=1}^{\infty}{\tilde q}_m(0|0,T)\,.
\end{equation}\\

\begin{figure}
\includegraphics[width=7.cm]{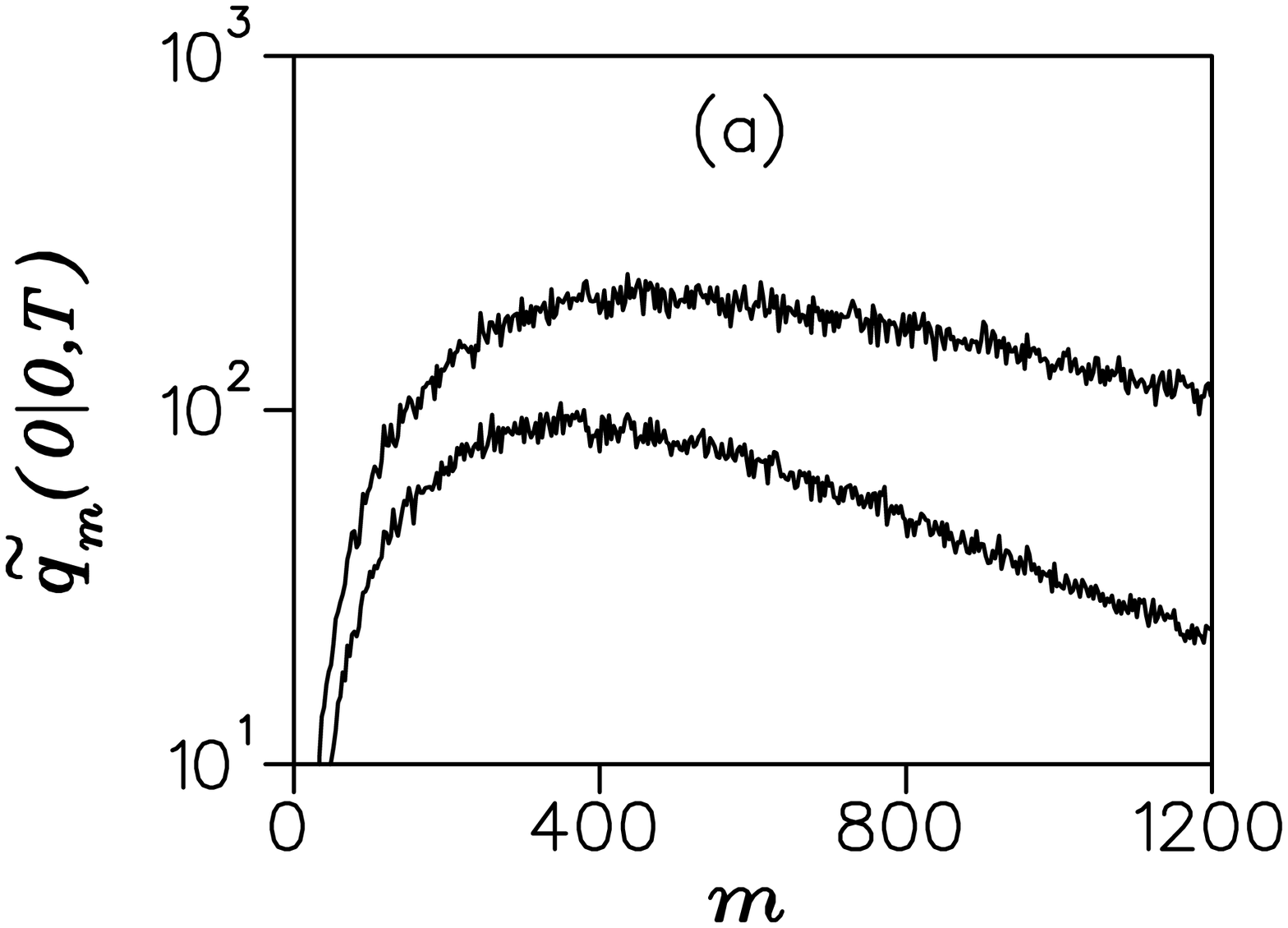}
\includegraphics[width=7.cm]{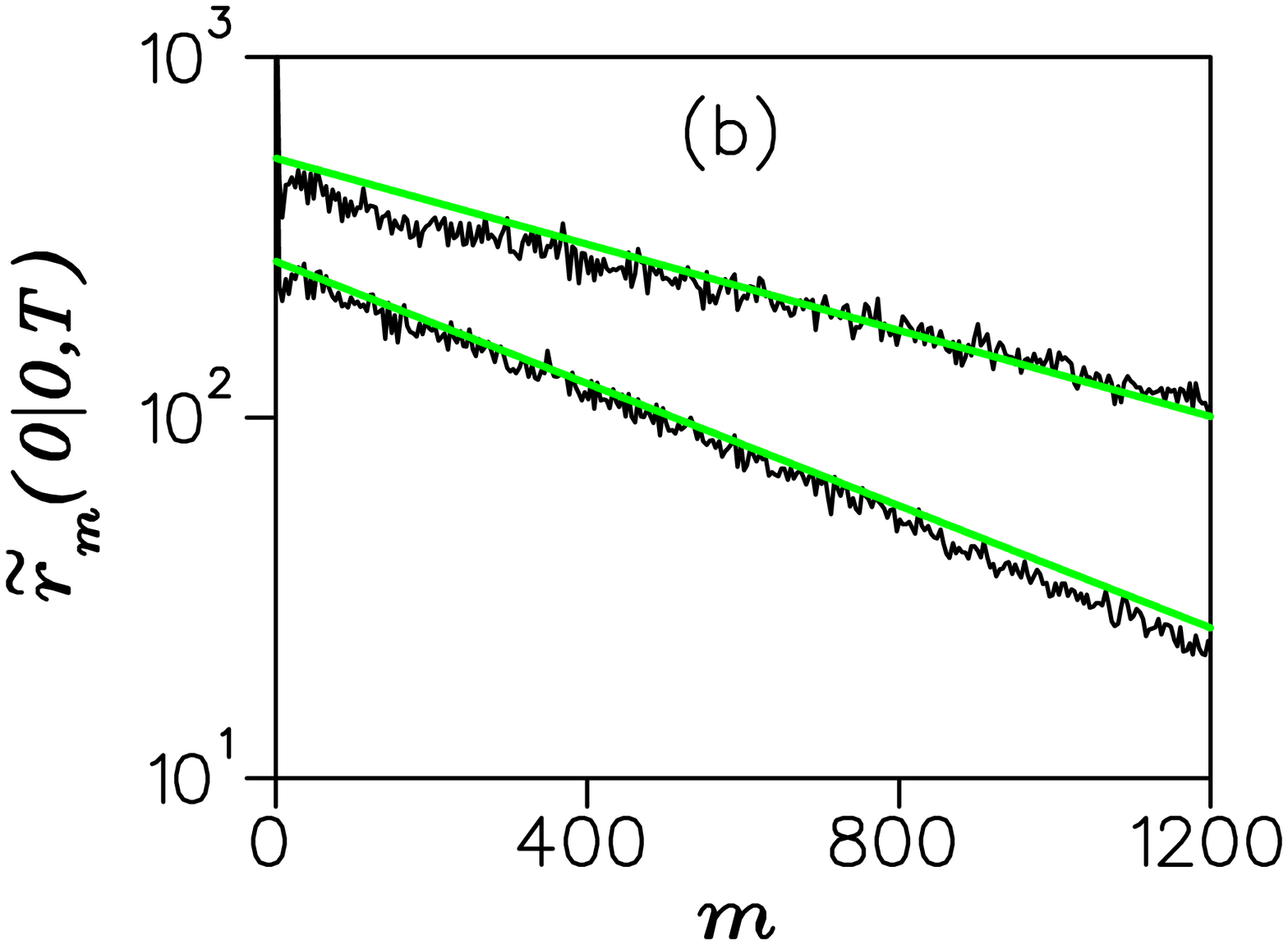}
\caption{(color online) Distribution ${\tilde q}_m$ (a) and
exponential decay of the factor $\tilde{r}_m$ (b), for
reversal times $T=50$ (bottom curves) and $T=100$ (top curves),
at $\hbar=1$, $g_0=2$, $\Delta=200$. Straight lines correspond
to $\exp(-2\Lambda m)$ [see Eq.~(\ref{muversusLambda})].}
\label{fig:Wdip}
\end{figure}

If the initial mixture is wide, $\Delta\gg \hbar$, the
pre-factor
$d_m\approx\left(1-e^{-\frac{\hbar}{\Delta}m}\right)^2$ is
small for all $m\lesssim \Delta/\hbar$. This explains the dip
seen in Fig.~\ref{fig:Wdip} (above). At the same time,
Fig.~\ref{fig:Wdip} (below) shows that the second factor
${\tilde r}_m(0|0,T)$ is very well described by an exponential.
Therefore we suppose that
\begin{equation}\label{r_mExp}
\begin{array}{c}
{\tilde q}_m(0|0,T)=\frac{{\tilde
q}_0(0|0,T)}{z(\mu)}\,d_m\,e^{-\mu(T)m}\,,\\
z(\mu)=\sum_{m=1}^{\infty}d_m\,e^{-\mu m}\,.
\end{array}
\end{equation}\\

To fix the two unknown functions ${\tilde q}_0(0|0,T)$ and
$\mu(T)$ we calculate the moments
\begin{equation}\label{Mean_m^k}
\begin{array}{c}
\tilde{\mathfrak m}^{(k)}(T)=\sum_{m=1}^{\infty}m^k {\tilde q}_m(0|0,T)=\\
={\tilde q}_0(0|0;T)
\,\frac{1}{z(\mu)}\, \left(-\frac{d}{d\mu}\right)^k\,z(\mu)\,.
\end{array}
\end{equation}
Then
\begin{equation}\label{Tildews}
\begin{array}{c}
{\tilde q}_0(0|0;T)=-\tilde{\mathfrak m}^{(1)}(T)
\frac{z(\mu)}{z'(\mu)}\,,\\
{\tilde q}_m(0|0;T)=-\frac{\tilde
{\mathfrak m}^{(1)}(T)}{z'(\mu)}\,d_m\,e^{-\mu m}\,,
\end{array}
\end{equation}
 where the prime denotes differentiation with respect to the parameter
$\mu$. Besides we obtain the connection
\begin{equation}\label{muversusLambda}
\Lambda(T)\equiv\frac{\tilde {\mathfrak m}^{(1)}(T)}{\tilde
{\mathfrak m}^{(2)}(T)} =-\frac{z'(\mu)}{z''(\mu)}\, ,
\end{equation}
which expresses the function $\mu(T)$ in the terms of the two
lowest momenta. Under the condition $\Lambda\ll 1$, which is
well justified by our numerical simulations, equation
(\ref{muversusLambda}) can be solved analytically resulting in
$\mu(T)\approx 2\Lambda(T)$. Thus the two lowest moments {\it
entirely fix} the probability distribution
$\tilde{\mathcal{W}}_{m}$.

\begin{figure}
\includegraphics[width=7.cm]{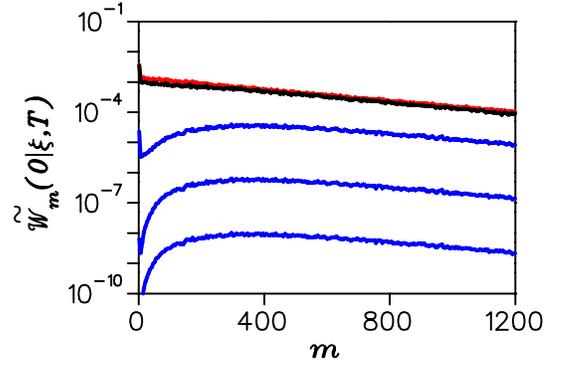}
\caption{(color online) Same as in Fig.~\ref{fig:expansatz2}
but starting from a mixed state with $\Delta=200$.}
\label{fig:expansatz3}
\end{figure}

When the ratio $\xi/\xi_c(T)$ exceeds the unity, the dip in the
probability distribution $\tilde{\mathcal{W}}_{m}(0|\,\xi;T)$
disappears and the following exponential fit works well for all
nonzero harmonics (see Fig.~\ref{fig:expansatz3}):
\begin{equation}\label{ExpRev_mProb}
\tilde{\mathcal{W}}_{m\geqslant
1}=\left(1-\tilde{\mathcal{W}}_{0}\right)\,(e^\mu-1)\, e^{-\mu m}\,.
\end{equation}
As above we can express ${\tilde{\mathcal{W}}}_{0}$ and $\mu$
in terms of two moments $\langle {\tilde
m}^k\rangle=\sum_{m=0}^{\infty} m^k {\tilde{\mathcal W}}_m$,
$k=1,2$. Now we easily find that
\begin{equation}\label{Rev_mProbFull}
\begin{array}{c}
\tilde{\mathcal{W}}_{0}=1-\langle {\tilde m}\rangle
\,\frac{2\Lambda}{1+\Lambda}\approx 1-2\langle {\tilde
m}\rangle\,\Lambda\,,\\
\tilde{\mathcal{W}}_{m\geqslant 1}=\langle {\tilde m}\rangle\,
\frac{4\Lambda^2}{1-\Lambda^2}\,
\left(\frac{1-\Lambda}{1+\Lambda}\right)^m\approx
4\langle{\tilde m}\rangle\,\Lambda^2\,e^{-2\Lambda m}
\end{array}
\end{equation}
where $\Lambda(T)=\frac{\langle {\tilde m}\rangle}{\langle{\tilde m}^2
\rangle}$.
A large number $\sim 1/2\Lambda(T)$ of harmonics have in this
case similar noticeable probabilities.

\begin{figure}
\includegraphics[width=7.cm,angle=0]{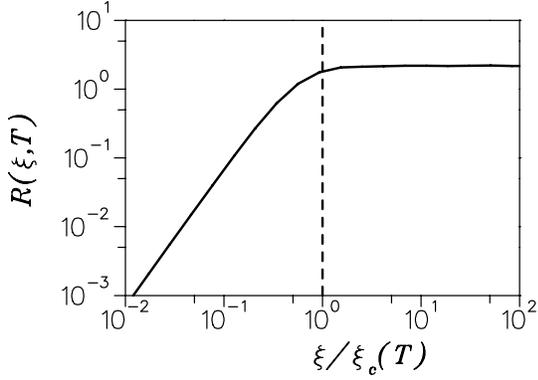}
\caption{Ratio $R(\xi,R)$ as a function of the parameter
$\xi/\xi_c(T)$, at $\hbar=2$, $g_0^2=6$, $\Delta=50$, and
$T=50$.} \label{fig:ratio}
\end{figure}

In Fig.~\ref{fig:ratio} the ratio $R(\xi,T)=\langle {\tilde
m(\xi,T)}^2 \rangle/\langle m^2\rangle_T$ is plotted as a
function of the parameter $\xi/\xi_c(T)$. When this parameter
is small the ratio is also small and proportional to $\xi^2$.
Since the fidelity $F(\xi;T)$ which describes the
redistribution of the excitation numbers after the backward
evolution is close to one under this condition, we conclude
that the initial state is recovered with good accuracy and the
motion is well reversible. On the contrary, when the
perturbation strength $\xi$ exceeds the critical value
$\xi_c(T)$ the fidelity becomes small and the residual number
of harmonics gets even larger than the number of harmonics
developed during the forward evolution. Therefore the evolution
becomes irreversible.
More precisely, for any reversal time $T$, there exists an
interval $0<\xi<\xi_c(T)$ of the perturbation strength $\xi$,
within which the quantum dynamics is approximately reversible.
This interval, is defined by the rate of proliferation of the
number of harmonics. In particular, this interval diminishes
exponentially fast when the semiclassical domain is approached.

\begin{figure}
\includegraphics[width=7.cm,angle=0]{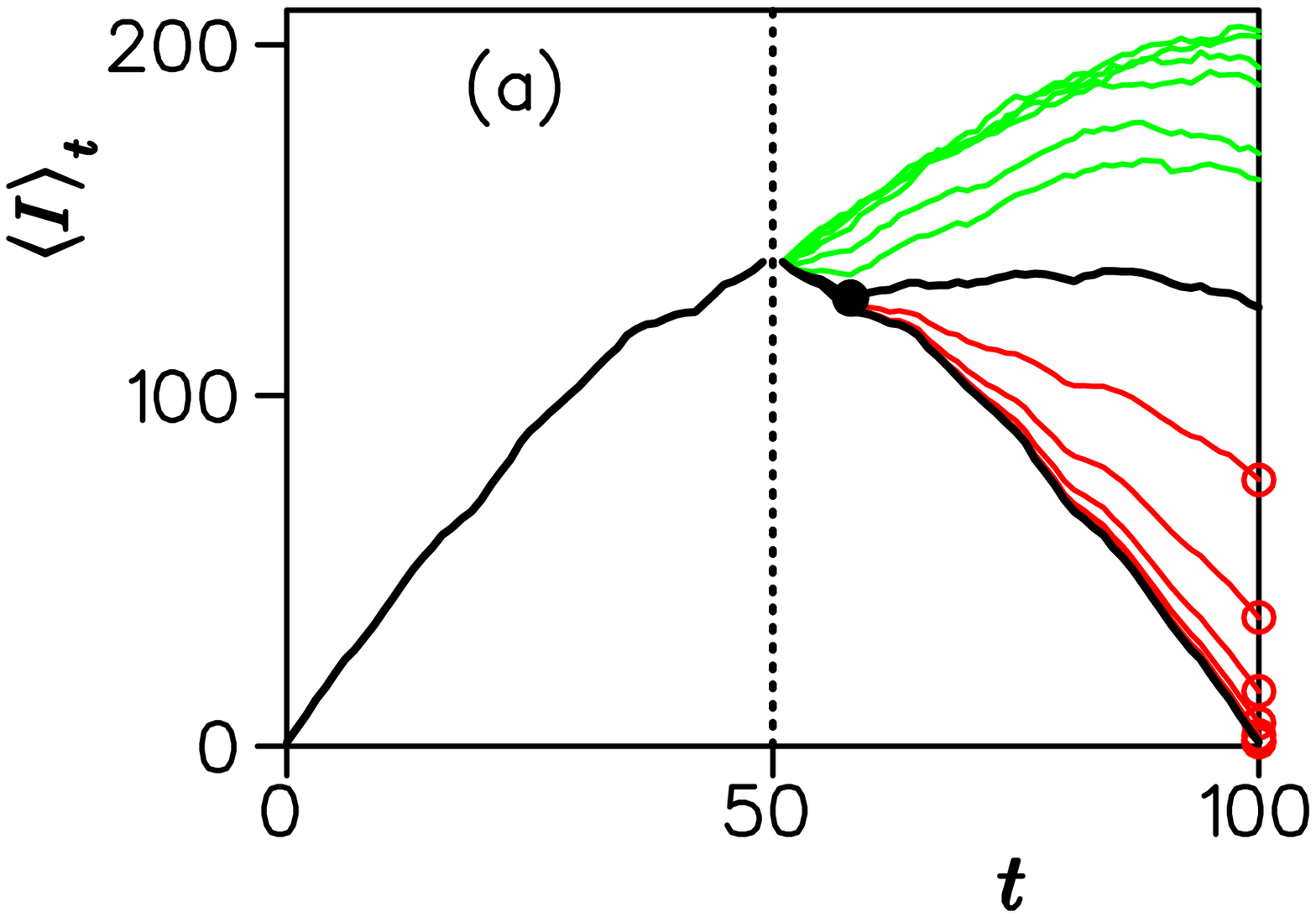}
\includegraphics[width=7.cm,angle=0]{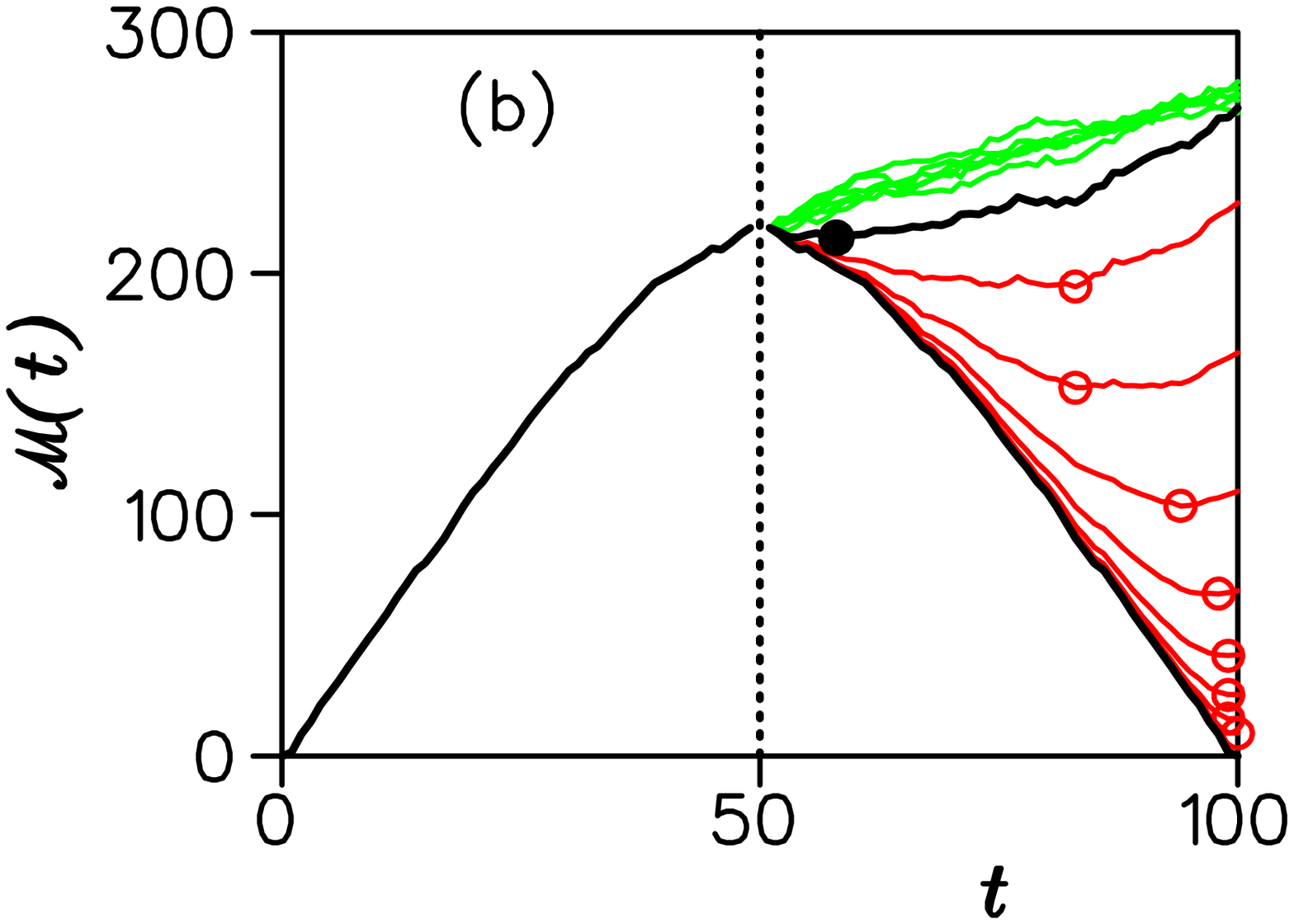}
\caption{(color online) Reversibility properties of quantum
dynamics, displayed by (a) $\langle I \rangle_t$  and (b) ${\cal
M}(t)=\sqrt{\langle m^2\rangle_t}$ , for different
values of the perturbation parameter: from bottom to top,
$\xi=\xi_c\times \exp{(l/2)}$, $l=-8,\ldots,-1$, $l=0$ (thick
black curve marked by the closed circle), and $l=1,\ldots,6$,
at $\hbar=1$, $g_0=2$, $\Delta=1$. Circles indicate positions
of the minima on the curves.} \label{fig:reversibility}
\end{figure}

At last the detailed temporal pattern is presented in
Figs.~\ref{fig:reversibility} of the backward evolution for
both the excitation number (panel (a)) as well as the number of
harmonics (panel (b)). It is clearly seen that the time
interval $\Delta t$ during which the system passes in reversed
order approximately the same sequence of the states, which it
does while evolving forward, decreases as a function of the
ratio $\xi/\xi_c(T)$. The existence of minimal deviation (or
the time of maximal return) during the backward evolution has
been stressed first in \cite{Kottos03, Kottos04}.

\section{Summary}

In this paper we have investigated the degree of stability and
reversibility of the quantum dynamics of classically chaotic
systems beyond the semi-classical domain. As a measure of
complexity we have used the number ${\cal M}(t)=\sqrt{\langle
m^2\rangle_t}$ of angular harmonics of the (initially
isotropic) Wigner function $W(I,\theta;t)$, developed during
the evolution for the time $t$. This number describes the
system's response to instantaneous rotation at that moment by
an infinitesimal angle $\xi\to 0$. The number ${\cal M}(t)$ is
found by calculating the distance between  the perturbed (rotated)
and unperturbed quantum distributions. We show that, in
contrast to the classical chaotic motion where the number of
harmonics grows exponentially, ${\cal M}_c(t)\sim e^{t/\tau_c}$
(with the rate $1/\tau_c$ increasing together with the Lyapunov
exponent), the number of harmonics of the quantum Wigner
function increases, after the Ehrenfest time, not faster than
linearly. This reveals much weaker sensitivity of the quantum
dynamics to perturbations than it is in the case of the
classical dynamics.

The relatively weak response of quantum systems to external
perturbations makes the quantum dynamics, to some extent,
reversible unlike the practically irreversible classical
chaotic dynamics. To quantify this statement we have analyzed
the degree of recovery of the initial, generally incoherent,
mixed state after the backward evolution of the quantum
distribution $W(I,\theta;t)$, rotated by a finite angle $\xi$
at some reversal moment of time $t=T$. The lack of the perfect
reversibility of the dynamics manifests itself by means of a
redistribution of the excitation numbers and by the number
${\tilde{\cal M}}(0)$ of harmonics of the Wigner function,
which remains after the backward evolution. Whereas the first
effect is directly described by the Peres fidelity the second
one is, in general, revealed with the help of an additional
infinitesimal rotation of the reversed state. We have shown
that there exists a critical value $\xi_c(T)=\sqrt{2}/{\cal
M}(T)$ of the perturbation strength $\xi$ such that the initial
state is well recovered if $\xi\ll\xi_c(T)$. Reversibility
disappears when the perturbation angle exceeds this value. The
interval of reversibility $0<\xi<\xi_c(T)$ exponentially
shrinks while approaching the semi-classical domain.

Thus our analysis establishes a direct quantitative connection
between the complexity of quantum phase-space distribution,
reduced in comparison to the classical dynamics, and the degree
of reversibility of the quantum dynamics.

\section*{Acknowledgements}

We acknowledge support by the Cariplo Foundation and INFN. V.S.
and O.Zh. acknowledge support by the RAS Joint scientific
program "Nonlinear dynamics and Solitons".


\begin{thebibliography}{01}

\bibitem{arrow}
D.L. Shepelyansky, Physica D {\bf 8}, 208 (1983); G. Casati, B.V.
Chirikov, I. Guarneri, and D.L. Shepelyansky, Phys. Rev. Lett. {\bf
56}, 2437 (1986).

\bibitem{chirikov}
B.V. Chirikov, F.M. Izrailev, and D.L. Shepelyansky, Sov. Sci. Rev.
C {\bf 2}, 209 (1981).

\bibitem{gu}
Y. Gu, Phys. Lett. A {\bf 149}, 95 (1990).

\bibitem{brumer}
A.K. Pattanayak and P. Brumer, Phys. Rev. E {\bf 56}, 5174 (1997);
J. Gong and P. Brumer, Phys. Rev. A {\bf 68}, 062103 (2003).

\bibitem{ikeda}
K.S. Ikeda, in {\it Quantum chaos: between order and disorder}, G.
Casati and B.V. Chirikov (Eds.) (Cambridge University Press, 1995).

\bibitem{peres}
A. Peres, Phys. Rev. A {\bf 30}, 1610 (1984).

\bibitem{fidelitynote}
Quantity (\ref{PFid}) has been widely
investigated in studies of the so-called quantum Loschmidt
echo~\cite{prosen}. However, we would like to
note that, similarly to Ref.~\cite{heller} and in contrast
to other previous studies~\cite{prosen},
the backward evolution proceeds with the same Hamiltonian as the
forward evolution and the perturbation acts instantaneously
only at the reversal time $T$.

\bibitem{prosen}
A review on the quantum Loschmidt echo is provided by
T. Gorin, T. Prosen, T.H. Seligman, and M. \v Znidari\v c,
Phys. Rep. {\bf 435}, 33 (2006). Ph. Jacquod and C. Petitjean,
preprint {\tt arXiv:0806.0987v1 [quant-ph] 5 Jun 2008}

\bibitem{heller}
C. Petitjean, D.V. Bevilaqua, E.J. Heller, and Ph. Jacquod,
Phys. Rev. Lett. {\bf 98}, 164101 (2007).

\bibitem{Glauber63} R.J. Glauber, Phys. Rev. {\bf 131} 2766 (1963).

\bibitem{Agarwal70} G.S. Agarwal and E. Wolf, Phys. Rev. D
{\bf 2}, 2161 (1970); {\it ibid.} 2187 (1970).

\bibitem{Schwinger53} J. Schwinger, Phys. Rev. {\bf 91}, 728 (1953).

\bibitem{Berman78} G.P. Berman, G.M. Zaslavsky, Physica A
{\bf 91}, 450 (1978); {\it ibid.} {\bf 97}, 367 (1979).

\bibitem{Sokolov84} V.V. Sokolov, {\it Nonlinear Resonance of
a Quantum oscillator} prep. Inst. Nucl. Phys. Siberia Div.
Acad. Sci USSR (1978); Teor. Mat. Fiz. {\bf 61}, 128 (1984) [Sov. J.
Theor. Math. {\bf 61}, 104 (1985)].

\bibitem{Sokolov07} V.V. Sokolov, G. Benenti,
and G. Casati, Phys. Rev. E {\bf 75}, 026213 (2007).

\bibitem{benenti02} G. Benenti and G. Casati,
Phys. Rev. E {\bf 65}, 066205 (2002); G. Benenti, G. Casati,
and G. Veble, Phys. Rev. E {\bf 67}, 055202(R) (2003).

\bibitem{Kottos03} T. Kottos and D. Cohen, Europhys Lett. {\bf 61},
431 (2003).

\bibitem{Kottos04} Moritz Hiller, Tsampikos Kottos, Doron Cohen and
Theo Geisel, Phys. Rev. Lett. {\bf 92} 010402 (2004).

\end{thebibliography}
\end{document}